\documentclass[onecolumnfleqn,usenatbib]{mnras}

\usepackage{newtxtext,newtxmath}

\usepackage[T1]{fontenc}
\usepackage{ae,aecompl}

\usepackage{graphicx}	
\usepackage{amsmath}	
\usepackage{gensymb}
\usepackage{booktabs}
\usepackage{multicol}
\usepackage{subfig}
\usepackage{float}

\title[Nothing to see here]{Nothing to see here: Failed supernovae are faint or rare}

\author[R. A. Byrne \& M. Fraser]{
R. A. Byrne,$^{1}$\thanks{E-mail: robert.byrne.6@ucdconnect.ie (RB)}
M. Fraser,$^{1}$
\\
$^{1}$School of Physics, University College Dublin, Belfield, Dublin 4, Ireland.\\
}

\date{Accepted XXX. Received YYY; in original form ZZZ}

\pubyear{2022}

\begin{document}
\label{firstpage}
\pagerange{\pageref{firstpage}--\pageref{lastpage}}
\maketitle

\begin{abstract}
The absence of Type IIP core-collapse supernovae arising from progenitors above 17 solar masses suggests the existence of another evolutionary path by which massive stars end their lives. The direct collapse of a stellar core to a black hole without the production of a bright, explosive transient is expected to produce a long-lived, dim, red transient known as a failed supernova. Despite the detection of a number of candidates for disappearing massive stars in recent years, conclusive observational evidence for failed supernovae remains elusive. A custom-built pipeline designed for the detection of faint transients is used to re-analyse 10 years of observations of 231 nearby galaxies from the PTF/ZTF surveys. This analysis recovers known supernovae, and yields a number of interesting transients. However, none of these are consistent with a failed supernova. Through Monte Carlo tests the recovery efficiency of our pipeline is quantified. By assuming failed supernovae occur as a Poissonian process with zero detections in the data set, 95 per cent upper limits to the rate of failed supernovae are calculated as a function of failed supernova absolute magnitude. We estimate failed supernovae to be less than 0.61, 0.33, 0.26, or 0.23 of the core-collapse SN rate for absolute magnitudes of $-11$, $-12$, $-13$, and $-14$ respectively. Finally, we show that if they exist, the Vera C. Rubin Observatory will find 1.7 -- 3.7 failed SNe per year for an absolute bolometric luminosity of $\sim 6 \times 10^{39} \textrm{ erg s}^{-1}$ out to distances of 33 - 43 Mpc, depending on their assumed spectral energy distribution.

\end{abstract}




\section{Introduction} \label{sec:introduction}

It is expected that most stars with masses above a threshold of $\sim$ 8 M$_\odot$ end their lives when the outward pressure provided by nuclear burning in their core can no longer support the star against the inward pull of gravity. The core collapses in on itself and generates an outward shock which destroys the star and produces a bright transient known as a core-collapse supernova (CCSN) \citep[e.g][]{2012ARNPS..62..407J}. The extreme physical conditions present at the onset of such a transient allow for the creation of heavy elements, making supernovae one of the most important classes of event for the study of galactic and chemical evolution \cite[e.g.][]{kobayashi2006}.

 For a number of nearby supernovae, it has been possible to directly identify the progenitor from archival data taken before the transient \cite[e.g.][]{2014MNRAS.439L..56F,Maund14,2017RSPTA.37560277V}. In doing so, the distribution of luminosities of core-collapse supernova progenitors can be studied. \cite{smartt2009} has noted that there is a lack of detected high-luminosity progenitors above a cutoff of $\log L / L_\odot \simeq 5.1$ dex. Comparing the observed distribution of CCSN progenitor masses to a Salpeter initial mass function reveals an inconsistency, implying that red supergiants (RSGs) above a mass of $\sim$ 17 M$_\odot$ may not produce bright CCSNe. On the other hand, \cite{daviesbeasor} have argued that the statistical significance of this apparent discrepancy is less than 2$\sigma$ once dust is accounted for and that no physical tension exists between the distributions of CCSN progenitors and red supergiants.

Some additional support for an absence of Type IIP SN progenitors above $\sim$ 17 M$_\odot$ comes from analysis of nebular spectra. The strength of emission lines from elements such as O and Mg is dependent on the mass of synthesised elements, and hence on the zero age main sequence mass of the progenitor. So far, high quality nebular spectra have been obtained for around two dozen nearby Type IIP SNe, and all have been found to be consistent with progenitors with mass $<$16 M$_\odot$ \citep{jerkstrand2015}.

One potential solution for the lack of CCSNe coming from massive RSGs is the possibility that the cores of larger stars are more likely to collapse directly to form a black hole without producing the outward shock required to detonate the remainder of the star, and thus not produce a bright transient that would allow astronomers to discover these events. \cite{pattonsukhbold} ran simulations evolving a grid of carbon-oxygen cores to the moment of core collapse in order to determine how their final structure, and thus their propensity to result in either an explosion or implosion depends on the initial mass of the core and its relative fractions of carbon and oxygen. It was found that the explodability of these cores varies non-monotonically, displaying "islands of explodability" where cores with certain physical characteristics are more likely to explode, causing a bright supernova, while others are more likely to implode, resulting in a failed supernova. Similar results, where discrete mass ranges of progenitors do or do not explode have also been found in other works \citep{oconnorott,2015ApJ...801...90P,ertl2016,muller2016,2016ApJ...821...38S,2020ApJ...890..127C}

A number of surveys have attempted to find instances of high-mass red supergiants disappearing without an associated supernova. \cite{reynolds2015} performed an analysis of archival data from the Hubble Space Telescope, consisting of 15 nearby galaxies. Within their data, they find one candidate consistent with a 25-30 M$_\odot$ yellow supergiant disappearing without a bright transient. A similar survey from \cite{kochanek2008} using data from the Large Binocular Telescope has found a $\sim$ 25 M$_\odot$ red supergiant candidate which underwent a weak optical outburst in 2009 before disappearing from view, with recent observations up until 2019 remaining consistent with the hypothesis that the star has disappeared \citep{adams2017,basinger2020}. A recent update from this survey has found a further candidate coming from the rapid disappearance of a blue supergiant in M101 \citep{neustadt2021}. However, each of these candidates cannot be conclusively confirmed as failed supernovae. For example, it is still possible that the progenitors for these events may still survive, but are obscured by enough dust to block them from detection. 

More recently, \cite{m51candidate} found another event in archival {\it HST} data of M51 which displayed possible signatures of a failed supernova event. In this case, a cool supergiant displayed optical variability before undergoing a dimming of over 2 magnitude at a point between 2017 and 2019. However, in 2021 the star rebrightened, leading \citeauthor{m51candidate} to reject the hypothesis of a failed supernova, instead interpreting this event as an analogue to the 2019-2020 Great Dimming of Betelgeuse \citep{betelgeuse}. These attempts at finding failed supernovae highlight a key difficulty with this method of discovery - such events may be difficult to distinguish from other phenomena, and unambiguously confirming their nature proves challenging.

While the simplest picture of a ``failed'' supernova is a massive star that simply disappears, it is still expected that there will be some optical emission associated with the collapse, albeit much fainter than in a classical core-collapse explosion.

\cite{nadezhin} first examined what the observational consequences of the loss of $\sim$ 1 M$_\odot$ from a $\sim$ 10 M$_\odot$ core as it collapses to a black hole would be. Such an event may not produce a shock strong enough to explode the progenitor star entirely, though it may result in the slow expulsion of the tenuously bound outer layers of the star. Following from this, \cite{lovegrove2013} studied the hydrodynamical response of a typical red supergiant to the sudden loss of 0.2-0.5 M$_\odot$ of gravitational mass from their cores due to neutrino emission at the point of core-collapse to a black hole. It is found that the expulsion of the outer layers of a red supergiant caused by this mass loss can produce a faint, long-lived, red transient. The total energy of such an event would be on the order of $10^{47}$ erg and would result in ejection velocities of $\sim$ 100 km s$^{-1}$ and luminosities of $\sim10^{39}$ erg s$^{-1}$ maintained for roughly a year. The detection of such a transient could provide direct evidence for a star which has collapsed to form a black hole without a bright supernova.

Such transients may be missed in surveys for a number of reasons. Firstly, they are intrinsically dimmer than traditional supernovae, and so their luminosities may lie around the detection threshold for many telescopes, making it more likely they will be missed. Additionally, since they are so long-lived, their evolution may span over multiple observing seasons, causing any variability to be missed. Finally, if these were not being actively searched for at the time, their low absolute magnitudes ($<-11$) could lead to them being discounted by teams explicitly searching for CCSNe.

In this paper we present a re-analysis of data from the Palomar Transient Factory \citep[PTF;][]{PTF} and its successor, the Zwicky Transient Facility \citep[ZTF;][]{ZTF}. Our data set covers 231 nearby galaxies, which we analyse with a custom pipeline, designed to search for long-lived faint transients. The combined PTF/ZTF data comprise roughly 10 years of observations with relatively regular cadence, particularly during the ZTF era. By reviewing these data, paying particular attention to high-quality stacking to produce deep images and following galaxies across multiple years we hope to maximise our chances of detecting long-lived faint transients. Additionally, through the injection of artificial failed supernova into our data sets, we quantify the detection efficiency of the pipeline for PTF/ZTF data, allowing us to place an upper limit on the rate of failed supernovae.

In Section \ref{sec:pipeline} we discuss the main elements of the pipeline used to analyse these data and search for faint transients. Section \ref{sec:eff} details the means by which we investigated the recovery efficiency of this pipeline through injection of artificial failed supernovae. In Section \ref{sec:sources} we present the sources which were found in our analysis. In Section \ref{sec:rates} we place an upper limit on the rate of failed supernovae, and compare this to that of core-collapse supernovae. In Section \ref{sec:discussion} we discuss the results we have obtained and their implications for our understanding of failed supernovae. Finally, in Section \ref{sec:conclusions} we summarise our conclusions from the analysis.

\section{Analysis Pipeline} \label{sec:pipeline}

We have developed a pipeline to search through our data set for potentially interesting transient sources. The pipeline reads in and reduces images for each galaxy under study, creating deep templates and stacks, before producing difference images for these. Persistent sources are found within these difference images, and photometry is carried out on each. 

The pipeline analyses the PTF and ZTF data separately. This is due to the two surveys using different filters - our PTF data consists of only images using the Mould-R filter, and our ZTF data uses the custom ZTF-r filter. Different filters are analysed separately to ensure consistency between templates and stacks, and thus increase the accuracy of our subtractions.

The main components of the pipeline are explained in detail below.

\begin{itemize}
    \item {\bf Galaxy sample}
    
    A list of galaxies to search for candidate failed SNe was obtained using an SQL query of the HyperLeda database\footnote{\url{http://leda.univ-lyon1.fr/}}. The goal was to find galaxies which would be visible from the Palomar Observatory, were close enough to make the detection of relatively faint transients feasible, and inclined at such an angle that galactic structure was visible so as not to miss transients due to galactic extinction.
     
     The constraints given were to return all galaxies at declination greater than -30$\degree$, with absolute magnitudes brighter than -18, inclined at less than 70$\degree$ to line-of-sight, and with a radial velocity corrected for Local Group infall onto Virgo of less than 1500 km s$^{-1}$. In total, 404 galaxies matching this description were found in the database.
     
    Any galaxy with fewer than 50 images in Mould-R for PTF, or ZTF-r for ZTF was discarded. This would ensure that analysis would only be carried out on galaxies with sufficient data to create a deep template and potentially find interesting transients. In total, we analysed PTF data for 217 galaxies, and ZTF data for 206 galaxies.

    \item {\bf Image preprocessing}
    
    The central coordinates, major and minor axes, and position angle of each galaxy in our sample are queried from HyperLeda \citep{hyperleda}. Each of the raw images is trimmed to a box centred on the galaxy with side length equal to the larger of 15 arcminutes or 1.5 times the galaxy's major axis.  This removes extraneous data, while keeping the images large enough to include the entire galaxy and sufficient background to find point sources for fitting PSFs, calibrating zero points, and performing accurate subtractions.
    
    In a small number of cases, a galaxy may lie close to the edge of the PTF or ZTF field of view. In these cases, the cutout region around the galaxy will be truncated. This is accounted for by scaling for the fraction of each galaxy that is observed when calculating detection efficiencies and rates (discussed later). These cases are not overly common in our data set, and we find that sufficient sources are still available for zero point calibration and fitting PSFs.

    Accurate alignment of images is necessary to produce deep stacks of images for templates or analysis, and later to ensure accurate image subtraction between the two.  This process is carried out using the Astroalign Python module which detects point sources in separate images, measures their pixel coordinates, derives geometric transforms between them, and remaps one image to another \citep{astroalign}.
    
    We examine the degree to whether alignment using Astroalign conserves flux. To do so, we generate 3000 images with 20 false PSF-like sources in each. Each image is rotated and blurred by a random amount. We then use Astroalign to align the original image to the offset image. Aperture photometry is performed on each source visible in the images both before and after alignment, and the fluxes of these sources are compared. From comparing the fluxes of $\sim$ 50000 sources, we find that the ratio is centred on 1, with a standard deviation of 0.016, corresponding to an error of 0.016 mag. This is small enough to be ignored compared to other sources of uncertainty.

    \item {\bf Building templates}
    
    The success of detecting particularly faint sources using difference imaging relies on the availability of a good quality template image to be subtracted from subsequent stacks of images. This template is created by combining multiple images taken under favourable conditions (i.e. deep, and good seeing) into a single master template.
    
    Three inputs are required to create such a deep template for a set of images: the range of time for the template to cover, an upper limit for the seeing of images to be used, and a lower limit on the zero point magnitude of images to be used. To find initial bounds for these cutoff points, a subset of PTF images was examined. The distributions of these parameters were plotted, and cutoff values were selected which would ensure that all template images were of a relatively high quality, but that enough images should be available. The upper bound for seeing was taken to be 2.2 arcseconds (i.e. only images with seeing lower than or equal to this value were included). The lower bound for the zero point was taken to be 22.9 (i.e. only images with zero points deeper than this were included). The same cutoffs were used for ZTF analysis. For both data sets, the maximum template length was set to 100 days.
    
    For each galaxy, the complete set of images is sorted by date of observation. Images are checked in order until one passes the specified quality thresholds. This is taken as the first image to be combined into the master template. All images prior to this are discarded. Any images taken within the template time range past this first image which also meet the quality thresholds are taken as template images. In the case where fewer than three images are eligible for the template, the cutoff points for seeing and image zero point are incrementally made more lenient until a sufficient number are obtained. Any images after the template cutoff point are left as science images.
    
    The template image with the best seeing is taken as the base, and all other template images are aligned to it using Astroalign. The images are trimmed to remove any edge defects caused by the alignment and the sigma-clipped median of all constituent template images is taken to produce a single master template.
    
    \item {\bf Building stacks of science images}
    
    Similarly to creating a deep template image, it is possible to increase the likelihood of detecting faint sources by producing stacks of science images. The pipeline attempts three separate types of stacking: monthly, weekly, and nightly. While it is possible to force the code to use a specific stacking length, in the absence of this, each length will be attempted in turn.
    
    The stacking algorithm attempts to use the longest stacking time possible in order to create the deepest possible stacks, while avoiding situations where so many images are included in a stack as to be wasteful. The first stack will include all images taken within 30 days of the first image.
    
    The next stack will begin with the first image after this first 30 day stacking window, and use all images within 30 days of that point. The process is repeated until all images are combined into stacks.
    
    An overstacking parameter can be specified, which corresponds to the maximum number of images permissible in a single stack. This is set arbitrarily at 15 images. The purpose of this parameter is to strike a balance between combining images into stacks in order to boost signal-to-noise ratio and preserving a good sampling cadence. The addition of further images to a stack can increase the stack's signal-to-noise ratio, but the effect that this has decreases with the number of images already present. Additionally, combining many images with drastically different seeing may have negative effects on the stack. For dimmer sources with marginal detections, it may also be more useful to have a larger number of stacks where the source can be detected, as opposed to a single one. For these reasons, we attempt to ensure overstacking does not occur.
    
    If any stack contains more images than this value during the process of monthly stacking, the code will instead apply weekly stacking. If overstacking persists, the code finally uses nightly stacking.
    
    Within each set of images comprising a single stack, the image with the best seeing is selected as the base. Each of the other images is aligned to this image, and the median of the images is taken to produce the stack. The master template is aligned to each stack in turn, creating an individual template image specific to each stack. This is done to minimise the amount of transformations applied to the stacks, which require greater accuracy in their photometry. The geometric transformations between the master template and each stack are saved in order to be able to easily compare coordinates between differently aligned stacks.
    
    In initial tests of the pipeline, we allowed the code to select its stacking duration using the above method. It was found that for the majority of galaxies, the pipeline elected to use nightly stacking over weekly or monthly. In addition to this, we performed preliminary runs examining the recovery efficiency of the pipeline (described further in the following section) while forcing stacking of each length. We found that, on average, nightly stacking produced the best results for recovering injected sources. This is presumably because the relatively small gain in sensitivity (which will increase as the square root of the total observation time) is negated by the effect of combining images with different seeing. For these reasons, we decided to force nightly stacking in our full analysis of the data sets. This allowed for consistent analysis across all galaxies.
    
    \item {\bf Astrometry}
    
    The master template for each galaxy is solved for accurate WCS information using astrometry.net \citep{astrometry.net}. As the transformations between the master template and each stack are saved, WCS information for the stacks can easily be derived without needing to run astrometry on all of them.
    
    \item {\bf Image subtraction}
    
    To create difference images the {\sc hotpants} code \citep{hotpants} is implemented. This package performs PSF matching and transformations between two aligned images and produces a subtraction of one image from another.
    
    {\sc hotpants} is used to subtract the individually aligned template images from their respective stacks. It was found that for these images, good quality subtractions were produced by using a spatial order of 1 for the kernel variation, a threshold for sigma clipping statistics of 5.0, a high sigma rejection for bad stamps in kernel fit of 4.0, an RMS threshold for good centroid in kernel fit of 10, and upper valid data counts of 40000 in the templates and stacks, with the produced difference images being normalised to the template. Images were divided into 5 $\times$ 5 regions of stamps and reference sources were chosen automatically. All other options were left at default.
    
    \item {\bf Building PSFs}
    
    PSFs are modelled for each of the stacks. This is necessary for a number of reasons. Firstly, it informs us of the shape of the sources in the images which will allow for more accurate source detection and aperture photometry. Secondly, a model PSF is required to be able to build artificial sources which will later be injected into images in order to test the efficiency of the pipeline.
    
    Code to build the PSF of the images was adapted from the AutoPHoT pipeline \citep{autophot}. This code fits 2D Gaussians to sources in each image in order to accurately determine the FWHM of sources in each dimension as well as the average residual deviation of the sources from a true Gaussian.
    
    \item {\bf Image calibration}
    
    Calibrating each of the images allows for accurate transformation of the fluxes measured from detected sources to their corresponding magnitudes, and is also necessary for the later injection of artificial sources. For each galaxy, a subset of the PTF Sources Catalog is downloaded from a region centred on the galaxy. To ensure accuracy in the calibration, only sources with a $>99$ per cent probability of being a star (\textsc{class\_star}) and with $<0.05$ mag uncertainty on magnitude (\textsc{magerr\_auto}) from the catalog are considered.
    
    Source detection is performed on each image, and aperture photometry is performed on each source to obtain a measurement of its flux. The coordinates of each source are compared with those from the catalog. Any source in the image whose coordinates match a source in the catalog to within half an arcsecond is considered a match. For each matching source, the flux is converted to an instrumental magnitude which is compared to the true magnitude from the catalog to determine a zero point. The sigma-clipped mean of these zero points is taken as the zero point for the image.
    
    \item {\bf Source detection}
    
    Source detection is carried out on the difference images to find changes between the template and stacks. This is carried out using the DAOStarFinder function in the Photutils package \citep{photutils}. The function detects sources with a FWHM specified by that of the original science image before subtraction which are $n \sigma$ above background levels. The FWHM of expected sources in each image is taken from the previously modelled PSFs.
    
    The majority of sources detected in difference images tend to be spurious. These can be caused by slight misalignments, bright or saturated sources, inaccuracies in subtraction, etc. We employ a number of methods to narrow down a list of all sources detected in all images to a smaller list of sources more likely to be of astrophysical origin than due to systematic issues.
    
    As we are interested in persistent astrophysical sources, we require that any source be detected at the same position in two or more consecutive images before it is analysed. This reduces spurious detections arising from issues with single stacks such as cosmic rays.
    
    Bright and saturated sources tend to cause inaccuracies in the difference images in their vicinity. This leads to a large number of spurious detections close to them. To screen these out, the coordinates of each detected source are examined in its corresponding template and stack. If the value of any pixel in a 23 $\times$ 23 cutout around the source exceeds 55000, this source is discarded.
    
    As we are primarily interested in transients happening within the galaxy, we also require detections to be located at a distance less than twice the major axis of the galaxy from its centre.
    
    While these cuts remove a large amount of spurious or irrelevant detections, the main bulk of accepting or rejecting sources for analysis is performed by a neural network implemented to classify sources as real or bogus.
    
    \item {\bf Real-bogus classification}
    
    In order to curate the large amount of sources detected down to a more manageable number of sources more likely to correspond to physical events, a supervised machine learning classifier was implemented into the pipeline. In order to train such a classifier, a sample of training and testing data for both spurious and real detections was required. 
    
    To produce a sample of spurious detections, a number of image subtractions were performed without the injection of any synthetic sources, and source detection was performed on the subtractions at $5\sigma$ significance. Cutouts at each detection were produced, corresponding to errors arising from inaccurate subtractions, saturated sources etc. As every detection was classified as bad, this sample may have included some true detections, but as the vast majority would be true negatives the sample should be sufficient.
    
    To generate a sample of good detections, a number of synthetic PSF-like sources were injected into images. These sources had apparent magnitudes randomly chosen between 17 and 22. Difference images were produced and cutouts were made at the positions of each of the injected sources.
    
    In total, 16592 sources were used for training and testing data. 8640 of these were injected sources, and the remaining 7952 were spurious detections. The data consisted of $23\times23$ pixel cutouts of difference images centred on the sources.
    
    We track the performance of our classifiers through their accuracy, where accuracy is defined as the fraction of all predictions that are correct.
    
    We began by training a simple random forest classifier to determine whether or not a cutout corresponded to a spurious or real source. 80 per cent of the data was partitioned for training, with the remaining 20 per cent was left for testing the accuracy of the trained model. Using a simple random forest model with 512 trees as implemented using the scikit-learn library for Python \citep{scikit}, an accuracy of 92.96 per cent was attained on the testing data.
    
    We found that we could further increase the accuracy of our classifications by using a convolutional neural network (CNN) instead of the random forest model. CNNs are commonly applied to problems of image classification or recognition due to their capabilities of learning structural patterns within images. Our CNN consisted of two convolutional layers, each containing 64 filters, and sweeping 3 $\times$ 3 pixel local receptive fields across the image. These were each followed by max-pooling layers checking 2 $\times$ 2 areas within their outputs. The output from the second max-pooling layer was passed to a final densely connected hidden layer with 128 neurons, which finally connected to the output layer consisting of a single neuron. 
    
    The convolutional layers and final hidden layer all used the rectified linear unit as their activation function, while the output layer used a sigmoid function in order to give a continuous output between 0 and 1, whereby 0 would correspond to a classification of a source as spurious, and 1 to classification as a real source. The CNN was trained using the Adam optimisation algorithm and made use of the binary cross-entropy loss function. The CNN was implemented using methods from the Keras API for TensorFlow in Python \citep{keras, tensorflow}.

    To maximise the effectiveness of this classifier, the most important metrics were the rates of True Positives and False Negatives. Specifically, optimisation of this classifier would have the True Positive Rate as high as possible and the False Negative rate as low as possible.
    
    To improve these metrics, the output threshold at which a source is classified as a positive detection can be lowered. To begin with, any sample with an output greater than 0.5 was treated as a good detection. At the cost of lowering the overall accuracy slightly, by increasing the rate of false positives and decreasing the rate of true negatives, we can improve our most important metrics.
    
    In order to quantifiably measure the effect of changing the threshold, a Receiver Operating Characteristic (ROC) curve was produced for the CNN, which is shown in Figure \ref{fig:ROC}. The threshold for classifying a source as real was varied between 0 and 1 in 0.01 increments.
    
    Most classifications made by the CNN lie very close to values of 0 or 1, with relatively few sources being classified at intermediate values. As such, shifting the threshold leads to only incremental changes in the accuracy as well as the numbers of true positives and false negatives. We decide to place the threshold for a good detection at 0.1. This keeps the accuracy high, ensuring that the CNN still rejects most true negatives, while only missing some more difficult true positives.
    
    \begin{figure}
        \centering
        \includegraphics[width=\linewidth]{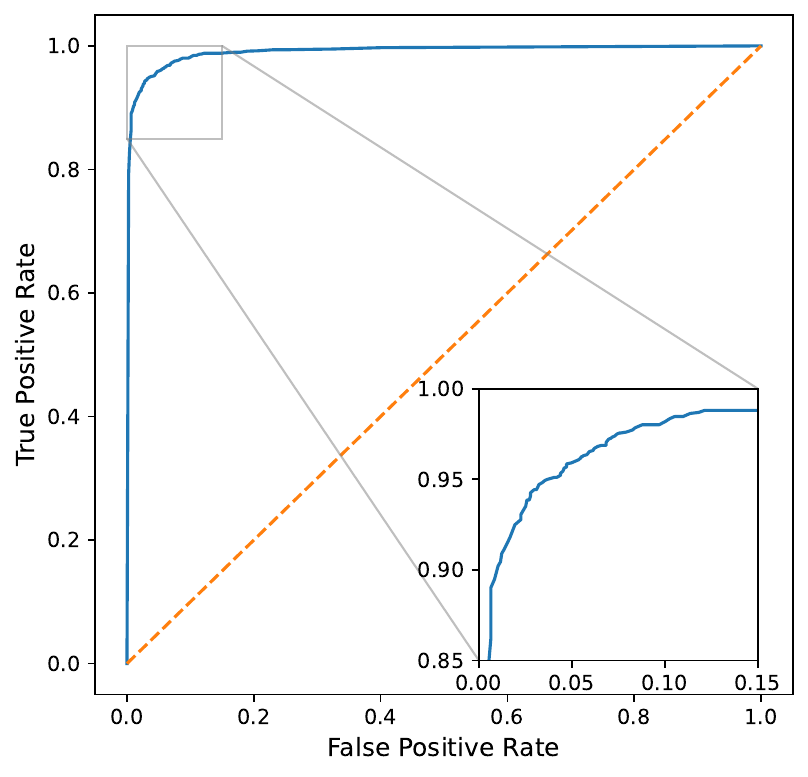}
        \caption{ROC curve for CNN designed to classify potential sources as real or bogus.}
        \label{fig:ROC}
    \end{figure}
    
    \item {\bf Human vetting}

    Three preliminary figures are produced for any sources passing all the criteria required to be classified as a good source.
    
    The first is a set of triptychs, displaying cutouts from the template, stacks, and associated difference images throughout the duration of the observation at the position of the detected source. These are useful for telling at a quick glance whether a source looks PSF-like, what its surroundings look like, whether it is close to any other sources etc.
    
    Next, light curves are produced across the entire observation in terms of flux and apparent magnitudes. These are produced for each image in the set regardless of whether the source was detected or not in an image. Forced photometry is carried out at the location, and upper limits for flux or magnitude are calculated for points where a source is not found. These light curves are useful for getting an idea of the speed at which a transient is evolving and its brightness, which can allow for predictions of the type of transient being observed.
    
\end{itemize}


\section{Recovery Efficiency} \label{sec:eff}

In order to place constraints on the rate of failed supernovae from the results obtained using the main pipeline, it was necessary to accurately determine the response of the pipeline to such events. This was done using a process of injecting artificial failed supernovae into the stacks for each galaxy and measuring how often these injected sources are recovered. The main steps by which this is carried out are detailed below.

\begin{itemize}
    \item {\bf Model for transients}
    
    A template light curve for a failed supernova is taken from \cite{lovegrove2013}. The light curve, which depicts the transient produced by a failed supernova from a red supergiant with a ZAMS mass of 15 M$_\odot$ lasting roughly 600 days was digitised.
    
    In order to test for different variations on this transient, the peak brightness and transient duration were left as customisable parameters. As such, the flux of the source could be scaled to investigate the effect this had on the ability to recover such sources.
    
    We chose to inject sources lasting 300 days. This was a conservative estimate, as the original model lasts for roughly 600 days. This means that our final calculated rates are based on transients which last for a shorter period of time than may be expected from true failed supernovae. By doing so, we can be certain that our upper limits are applicable to longer transients, as they are calculated on transients which should technically be more difficult to detect. However, it is likely that this choice does not affect efficiency majorly, as in either case the transient spends more than an observing season at roughly constant brightness.
    
    The model light curve we use peaks at a bolometric luminosity of $\sim 6 \times 10^{39} \textrm{ erg s}^{-1}$, corresponding to a bolometric luminosity of $\sim-10.75$. The magnitude of such an event in individual filters would depend on its colour. While it would be interesting to attempt injection of sources fainter than this, they would not be detectable in the majority of data from PTF or ZTF. For this reason, we choose to inject sources at absolute magnitudes of -11, -12, -13, and -14. 
    
    Figure \ref{fig:lims} displays the limiting magnitudes for ZTF observations of NGC 4435 compared to injected Lovegrove and Woosley-like light curves for our chosen magnitudes, each lasting 300 days.For the brighter sources, it can be seen that the peak brightness exceeds the majority of limiting magnitudes. However, for our dimmest sources at absolute magnitudes of -11, we begin to see images with limiting magnitudes above our peak brightness. We hence do not inject sources fainter than magnitude -11. We discuss this point further in Section \ref{sec:discussion}.
    
    \begin{figure}
        \centering
        \includegraphics[width = \linewidth]{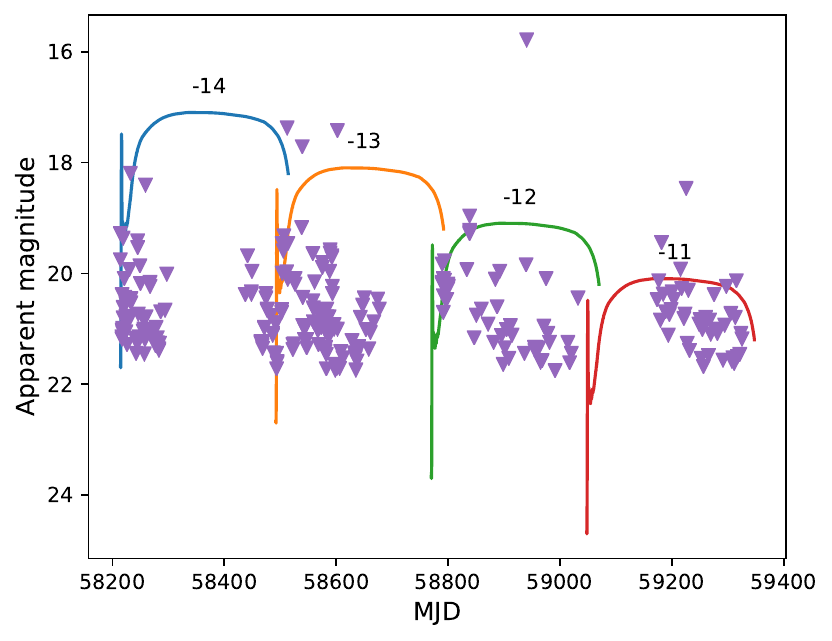}
        \caption{Comparison of limiting magnitudes for ZTF observations of NGC 4435, marked by purple triangles, with light curves of Lovegrove and Woosley-like failed supernovae each lasting 300 days and peaking at magnitudes -11, -12, -13, and -14. NGC 4435 is chosen as a representative galaxy for our sample as its distance modulus of 31.09 is the closest to the mean distance modulus of the entire sample, 31.1. Text above light curves displays the absolute magnitude of each source.}
        \label{fig:lims}
    \end{figure}
    
    \item {\bf Choosing injection sites}
    
    In order to find the most likely positions for a failed supernova to occur, we track star formation in the galaxy by assigning probabilities of being chosen to each pixel in an image based on the fraction of galaxy light at these locations. For this, we used R-band images from the ESO Online Digitized Sky Survey\footnote{\url{https://archive.eso.org/dss/dss}}. This survey was complete for all galaxies in our sample, giving us access to full images of each galaxy.
   
   An elliptical region is defined within each image from the DSS-2-red survey corresponding to the spatial extent of the galaxy. The WCS coordinates and flux of each pixel within this region are saved. The fluxes are normalised such that their cumulative sum equals one. The fluxes and their corresponding coordinates are sorted in order of ascending flux. The idea for this analysis is similar to that of \cite{ja06}, wherein the correlation between the positions of Type II SNe and sites of H$\alpha$ emission was examined.
   
   To choose a position at which to inject a source, a random number between 0 and 1 is generated. The coordinates corresponding to the first cumulative level of flux exceeding this threshold is chosen as the position at which to inject the source. Due to the monotonically increasing derivative of the cumulative fluxes, this means that brighter regions will be preferentially chosen as transient sites over dimmer regions.
   
   \item {\bf Choosing injection times}
   
   The times at which injected sources occur are chosen randomly. For each source, this is done by picking any MJD between that of the first science image and the final one. This corresponds to the beginning of the template light curve. Interpolation of the light curve is performed to calculate the expected magnitude of the failed supernova in each of the stacks following this time. This process is carried out separately for each source in each galaxy, so all sources will begin at different times.
    
    \item {\bf Building and injecting sources}
    
    Building and injecting realistic artificial sources requires a PSF to be fit to each image during the running of the main pipeline, as well as accurate calibration of the zero point. From the randomly selected starting time for the transient and the interpolated expected magnitudes for the following images, the PSF can be scaled to a flux corresponding to the correct magnitude source in each image. For any images prior to injection of the transient or after its conclusion, the PSF is simply set to zero. Values from the scaled PSFs are added to each image at the injection site.
    
    It was tested whether multiple artificial sources could be independently injected at one time without impacting each other. This would allow for larger quantities of sources to be injected in single runs, speeding up our estimations of recovery efficiency. One way in which this could alter results would be by lowering the overall quality of subtractions.
    
    As subtraction with {\sc hotpants} relies on measuring the PSF of sources in the template and science images, it was thought that injecting a large number of false sources into one image might negatively impact the accuracy at which the subtractions were made. To test this, 200 artificial sources were injected into one image, and a subtraction was produced. It was found that even with this large number of additional sources, the quality of the subtraction was very similar to a control subtraction with no additional sources. It was thus concluded that the number of injected sources was of little concern to subtraction quality.
    
    Another issue could occur if injecting too many sources resulted in two being placed coinciding with one another, or close enough to disrupt accurate measurement of either PSF. This is of particular importance for smaller galaxies where there are fewer positions for sources to be injected, and thus much higher chance of this situation.
    
\end{itemize}

For each galaxy, a maximum of 10 sources are injected. To do so, a position and time of injection are generated using the methods described in the previous section. As some images do not include the entirety of the galaxy, it is possible that the image will not contain this position. If this occurs, a source is not injected. However, this source is still counted for calculation of recovery efficiencies, as these are reasonable positions for such a source to occur which would be missed. This choice is accounted for when determining the final limits on the failed supernova rate by factoring in the fraction of the galaxy's light visible in the image during our final calculations.

As well as this, it is also possible that a source will be injected during a period between observing seasons where no images of a galaxy are available. No attempt is made to inject sources preferentially during periods where imaging is available, as it is reasonable to assume a failed supernova could occur at any point and this will impact the efficiency of the survey.

At every attempt to inject a source, the code first checks that no other source has been injected within 20 arcseconds of the new candidate. If another source is present, the new source is not injected. In doing this, galaxies with larger angular sizes can contain a large number of sources while smaller galaxies have only what they can handle without overcrowding.

The total numbers of simulated sources of each magnitude injected into images for each survey are presented in Table \ref{tab:injected}.

\begin{table}
    \centering
    \resizebox{\linewidth}{!}{%
    \begin{tabular}{@{}ccc@{}}
        \toprule
        Source Magnitude & Sources injected in PTF data & Sources injected in ZTF data   \\ \midrule
        -11              & 7744                         & 7555                          \\
        -12              & 7780                         & 7509                          \\
        -13              & 7724                         & 7503                          \\
        -14              & 7806                         & 7529                          \\ \bottomrule
    \end{tabular}%
    }
    \caption{Total numbers of simulated sources injected into PTF and ZTF data sets for calculation of recovery efficiency. For the 217 galaxies analysed using PTF data, this results in an average of 36 injected sources of each magnitude per galaxy. For the 206 galaxies analysed using ZTF data, this results in an average of 37 injected sources of each magnitude per galaxy.}
    \label{tab:injected}
\end{table}

The stacks with injected sources are analysed using the subtraction and source detection elements from the main pipeline, and the coordinates of found sources were compared with those injected to determine the fraction of injected sources the pipeline typically recovered. Sources were injected at absolute magnitudes of $-14$, $-13$, $-12$, and $-11$, with each transient lasting 300 days. These were converted to apparent magnitudes for each galaxy using its distance modulus taken from HyperLeda.

To investigate the effect of reddening, we took a subset of 50 of the galaxies in our overall sample and examined them. We took vales for their foreground galactic extinction from the NASA Extragalactic Database (NED; \footnote{\url{https://ned.ipac.caltech.edu/}}). It was found that their mean extinction in the R-band was 0.088 mag. This difference was taken as low enough to not affect the calculated recovery efficiencies significantly. As such, we neglect reddening.

The injection and source detection processes were carried out 5 times for sources of each magnitude for each galaxy.

\begin{figure}
    \centering
    \includegraphics[width = \linewidth]{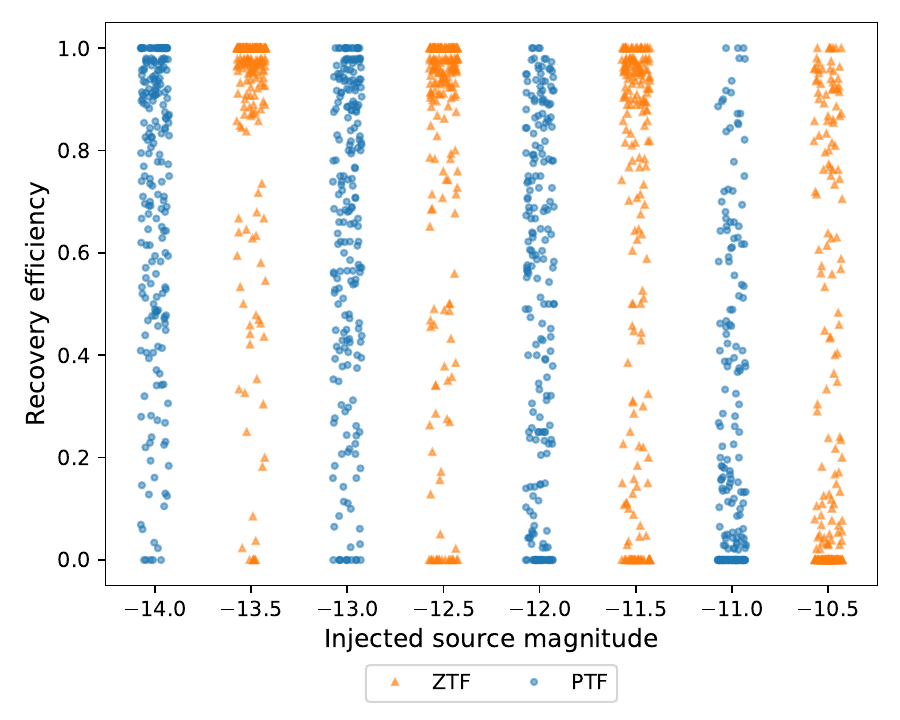}
    \caption{Efficiency of pipeline at recovering injected failed supernovae at absolute magnitudes of $-14$, $-13$, $-12$, and $-11$ from PTF/ZTF data sets. Each point shows the recovery efficiency for a single galaxy for either PTF or ZTF. For display purposes, ZTF efficiencies have been offset by 0.5 mag: for example, the cluster of orange triangles above magnitude -13.5 refers to the recovery efficiency for sources of magnitude -14 in ZTF data. Additionally, efficiencies are offset from their central magnitude by a random value of up to 0.075 mag in either direction in order to better display the overall spread of efficiencies. Typical error bars for ZTF are 0.06~mag, while typical error bars for PTF are 0.1~mag.}
    \label{fig:eff_mag}
\end{figure}

We calculated the recovery efficiency for sources of each magnitude for each galaxy by comparing the number of sources recovered with the number injected over the 5 iterations. The uncertainty on this efficiency was taken as the standard deviation of the efficiencies calculated by each individual iteration. As we had forced nightly stacking for our analysis of the data previously, we forced nightly stacking again so that our efficiency calculations direct relate to our initial search. The recovery efficiencies for sources of each magnitude for each galaxy using both PTF and ZTF data are displayed in Figure \ref{fig:eff_mag}.

For both sets of data, a large amount of galaxies have efficiencies clustering around either 0 or 1. This implies that for many galaxies, if one source is recoverable, it is likely that most sources will be recoverable, and conversely if one source is missed, it is likely that most sources will not be detectable. This effect is most visible in the ZTF data set, while the PTF efficiencies do show a considerable amount of intermediate values. The reason for this is likely due to the improved cadence of ZTF compared to PTF. ZTF's more regular imaging allows for the detection of most sources in galaxies which are near enough for an injected source to be visible. Some potentially visible sources may be missed in PTF data due to large gaps in observation.

Mean values for the efficiencies are calculated for each magnitude, which are displayed in Table \ref{tab:effs}.

\begin{table}
    \centering
    \resizebox{\linewidth}{!}{%
    \begin{tabular}{@{}ccc@{}}
        \toprule
        Source Magnitude & Mean PTF Efficiency & Mean ZTF Efficiency \\ \midrule
        -11              & (24 $\pm$ 7)\%      & (33 $\pm$ 6)\% \\
        -12              & (50 $\pm$ 11)\%     & (64 $\pm$ 7)\% \\
        -13              & (64 $\pm$ 11)\%     & (78 $\pm$ 6)\% \\
        -14              & (71 $\pm$ 10)\%     & (87 $\pm$ 5)\% \\ \bottomrule
    \end{tabular}%
    }
    \caption{Mean recovery efficiencies for sources of different magnitudes from PTF and ZTF data. (87 $\pm$ 5) per cent of all injected Lovegrove \& Woosley-like sources peaking at magnitude -14 are recovered from ZTF data, including those occurring when a galaxy is not under observation.}
    \label{tab:effs}
\end{table}

We find that under these conditions, the pipeline should recover 71 per cent of failed supernovae present in analysed PTF data, and 87 per cent of failed supernovae present in ZTF data, assuming transients peaking at absolute magnitude $-14$. As the peak of the injected transient dims, these recovery efficiencies decrease similarly to 64/78 per cent, 50/64 per cent, and 24/33 per cent.

We see that ZTF data yields a higher overall recovery efficiency compared to PTF. This is to be expected, as ZTF images are deeper, allowing for the detection of fainter transients, as well as having a higher and more regular cadence, ensuring that there are fewer opportunities for transients to occur during a period where no observations are made of its host.

To examine the factors contributing to our recovery efficiencies, a sample of 13 PTF galaxies with efficiencies less than 100 per cent were selected at random. Each injected source which was not recovered was examined in further detail. This was done to determine whether the non-recovery of sources was attributable to the pipeline itself, or intrinsic to the data. A total of 37 missed injected sources were examined.

\begin{itemize}
    \item 29 sources were missed due to the timing of their injection. The injected sources last for 300 days before fading from view. It was found that, due to the observing cycles of PTF, a sizable number of sources were injected during periods where no images were available for the galaxy in question, leaving no possibility of source recovery. This demonstrates one of the main reasons why ZTF efficiencies are higher: ZTF's cadence is far more regular and has much fewer long gaps in observation compared to PTF.
    
    \item 3 sources were missed when the position generated for them lay outside the frame visible in the PTF images. For the majority of galaxies, their full extent is visible in observations, but a small number have sections cut off. These injections were counted as they represent reasonable locations where a source may appear but also have no possibility for being recovered with the images available.
    
    \item 3 sources were missed due to coincidence with saturated sources. As saturated sources tend to account for a substantial amount of false detections due to the difficulty in performing accurate image subtractions in their vicinity, sources detected close to them are dismissed.
    
    \item 2 sources were missed for reasons unrelated to image cadence, geometry, or source screening. In each of these cases, both in separate galaxies, the source was visible in a total of 3 images. In each of these images, the source was particularly faint. These sources count as the only sources in the sample which could reasonably be expected to be recovered based on the criteria within the pipeline.
\end{itemize}

\section{Sources Found} \label{sec:sources}

Interesting sources from the analysed PTF/ZTF data sets were selected by manually looking through the triptychs and preliminary light curves generated for found sources in each galaxy. Even with the implementation of real-bogus classification using a neural network, the majority of these sources still came from seemingly spurious detections. Sources were classified as interesting if there was a PSF-like source visible in the difference images, preferably with no obvious progenitor visible in the template so as to exclude variable stars in favour of new transients. As such, the lists of interesting sources presented below in Tables \ref{tab:PTFsources} and \ref{tab:ZTFsources} do not represent the totality of sources found in the data, simply the more interesting ones and those most likely to come from transient events.

After choosing these sources, more careful and thorough photometry was performed on each to produce the most accurate light curves possible.

The PTF sources include six previously known supernovae, one previously known supernova impostor, one long-period variable, one known asteroid, one known nova, and one unknown source warranting further analysis. The ZTF sources include three previously known supernovae, one supernova candidate, three long-period variables, three known luminous blue variables (LBV), one known intermediate luminosity red transient (ILRT), two known luminous red novae (LRNe), one known Seyfert galaxy, two known novae, and two unknown sources warranting further analysis.

\begin{table*}
    \centering
    \begin{tabular}{cccccc}
        \hline
        Name             & RA        & Dec          & Peak Magnitude   & Host Distance Modulus & Description                     \\\hline
        NGC0428OT2013-01 & 18.22915  & $+0.97979$   & 16.69            & 30.87                 & Known SN SN2013ct               \\
        NGC0598OT2010-01 & 23.40356  & $+30.77293$  & 17.80            & 24.71                 & Known nova M33 2010-07a         \\
        NGC0598OT2010-02 & 23.55901  & $+30.57162$  & 16.46            & 24.71                 & Long period variable            \\
        NGC0918OT2011-01 & 36.45357  & $+18.53319$  & 14.75            & 31.14                 & Known SN SN2011ek               \\
        NGC1068OT2011-01 & 40.67253  & $-0.00579$   & 13.30            & 30.12                 & Known asteroid 735 Marghanna    \\
        NGC3166OT2012-01 & 153.44987 & $+3.43401$   & 17.05            & 31.45                 & Known SN SN2012cw               \\
        NGC3344OT2012-01 & 160.89192 & $+24.89139$  & 18.65            & 29.96                 & Known SN SN2012fh               \\
        NGC4258OT2010-01 & 184.71855 & $+47.30790$  & 19.77            & 29.41                 & Unknown candidate               \\
        NGC5806OT2012-01 & 224.99625 & $+1.88993$   & 16.00            & 31.58                 & Known SN SN2012p                \\
        NGC5806OT2013-01 & 225.00073 & $+1.88142$   & 16.97            & 31.58                 & Known SN iPTF13bvn              \\
        NGC5806OT2014-01 & 224.99782 & $+1.90732$   & 18.31            & 31.58                 & Known SN impostor SNHunt248     \\\hline

    \end{tabular}
    \caption{Interesting sources found in PTF data. Distance moduli taken from HyperLeda \citep{hyperleda}. Reported peak magnitudes are observed values, uncorrected for distance or extinction.}
    \label{tab:PTFsources}
\end{table*}

\begin{table*}
    \centering
    \begin{tabular}{cccccc}
        \hline
        Name              & RA        & Dec         & Peak Magnitude   & Host Distance Modulus & Description                            \\\hline
        NGC0598OT2018-01  & 23.36387  & $+30.69074$ & 18.64            & 24.71                 & Long period variable                   \\
        NGC0598OT2018-02  & 23.41439  & $+30.72137$ & 18.71            & 24.71                 & Long period variable                   \\
        NGC0598OT2018-03  & 23.56712  & $+30.63379$ & 19.17            & 24.71                 & Long period variable                   \\
        NGC0598OT2019-01  & 23.48714  & $30.54269$  & 17.52            & 24.71                 & Known nova AT2019gc                    \\
        NGC1068OT2018-01  & 40.67206  & $-0.0088$   & 14.93            & 30.12                 & Known SN SN2018ivc                     \\
        NGC1385OT2020-01  & 54.37473  & $-24.501$   & 19.88            & 30.20                 & Known LBV AT2020pju                    \\
        NGC3031OT2018-01  & 148.96882 & $+69.03357$ & 19.67            & 27.78                 & Known nova M81 2018-11a                \\
        NGC3310OT2021-01  & 159.69692 & $+53.50854$ & 15.59            & 31.23                 & Known SN SN2021gmj                     \\
        NGC3423OT2019-01  & 162.79921 & $+5.84229$  & 17.64            & 30.40                 & Known LBV AT2019ahd                    \\
        NGC3729OT2018-01  & 173.46639 & $+53.11876$ & 18.24            & 31.58                 & Known LRN AT2018hso                    \\
        NGC4559OT2018-01  & 188.96789 & $+27.93216$ & 16.70            & 29.97                 & Known LBV AT2016blu                    \\
        NGC4725OT2019-01  & 192.68282 & $+25.60711$ & 19.31            & 30.41                 & Background SN candidate                \\
        NGC4826OT2019-01  & 194.12059 & $+21.70219$ & 19.96            & 28.22                 & Unknown candidate                      \\
        NGC5068OT2020-01  & 199.75808 & $-21.05455$ & 17.78            & 28.56                 & Known LRN AT2020hat                    \\
        NGC5194OT2019-01  & 202.42715 & $+47.18806$ & 16.53            & 29.67                 & Known ILRT AT2019abn                   \\
        NGC5194OT2020-01  & 202.47206 & $+47.22349$ & 19.35            & 29.67                 & Unknown candidate                      \\
        NGC5457OT2020-01  & 211.07004 & $+54.27087$ & 19.90            & 29.26                 & Seyfert galaxy 3XMM J140416.7+541615   \\
        UGC06930OT2020-01 & 179.31124 & $+49.29223$ & 14.46            & 30.87                 & Known SN SN2020rcq                     \\\hline

    \end{tabular}
    \caption{Interesting sources found in ZTF data. Distance moduli taken from HyperLeda \citep{hyperleda}. Reported peak magnitudes are observed values, uncorrected for distance or extinction.}
    \label{tab:ZTFsources}
\end{table*}

\subsection{Known Sources}

\begin{figure*}
    \centering
    \subfloat[]{
      \includegraphics[width=0.49\linewidth]{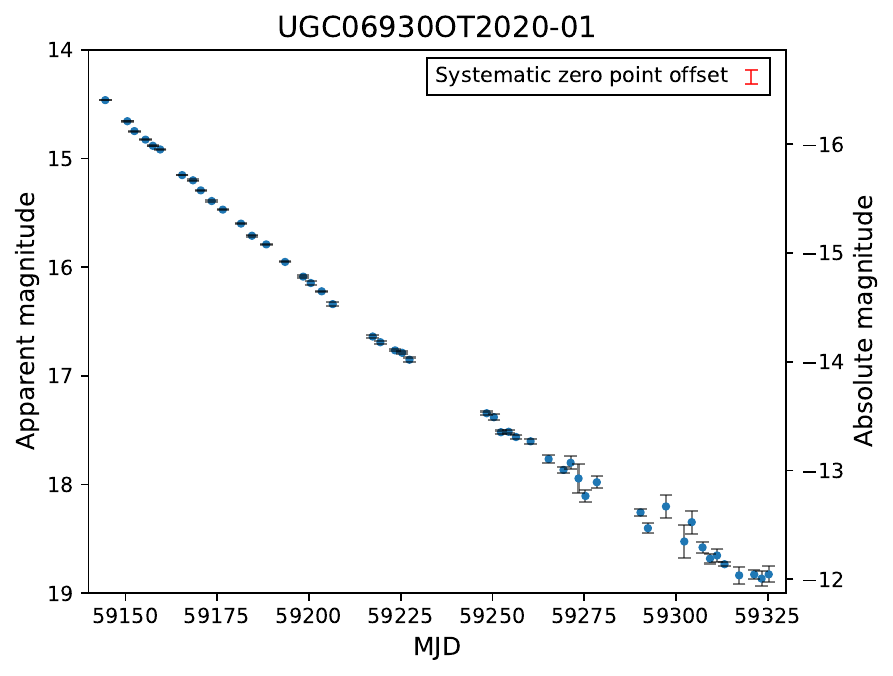}
    }
    \subfloat[]{
      \includegraphics[width=0.49\linewidth]{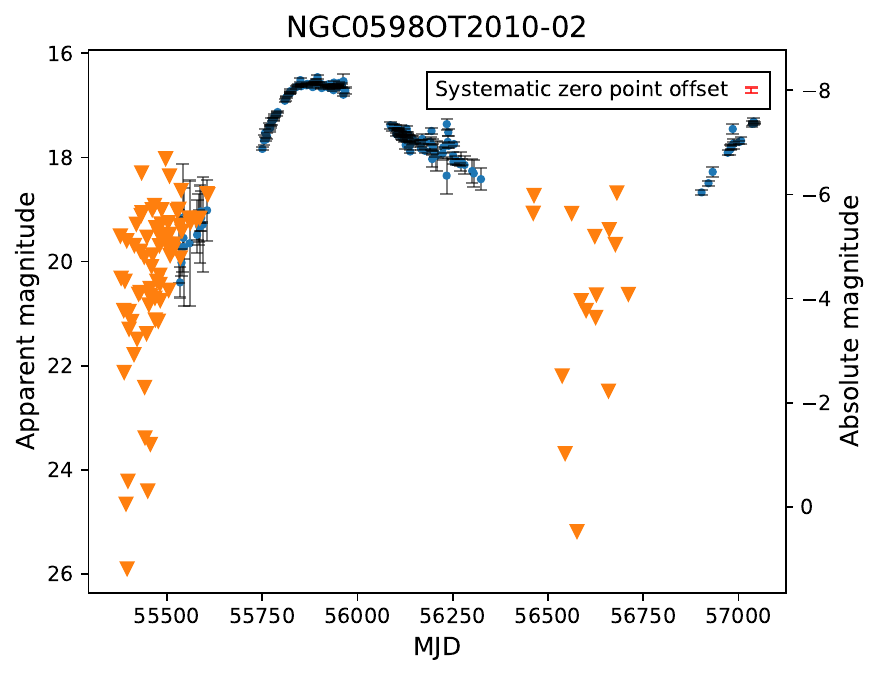}
    }
    
    \hspace{0mm}
    
    \subfloat[]{
      \includegraphics[width=0.49\linewidth]{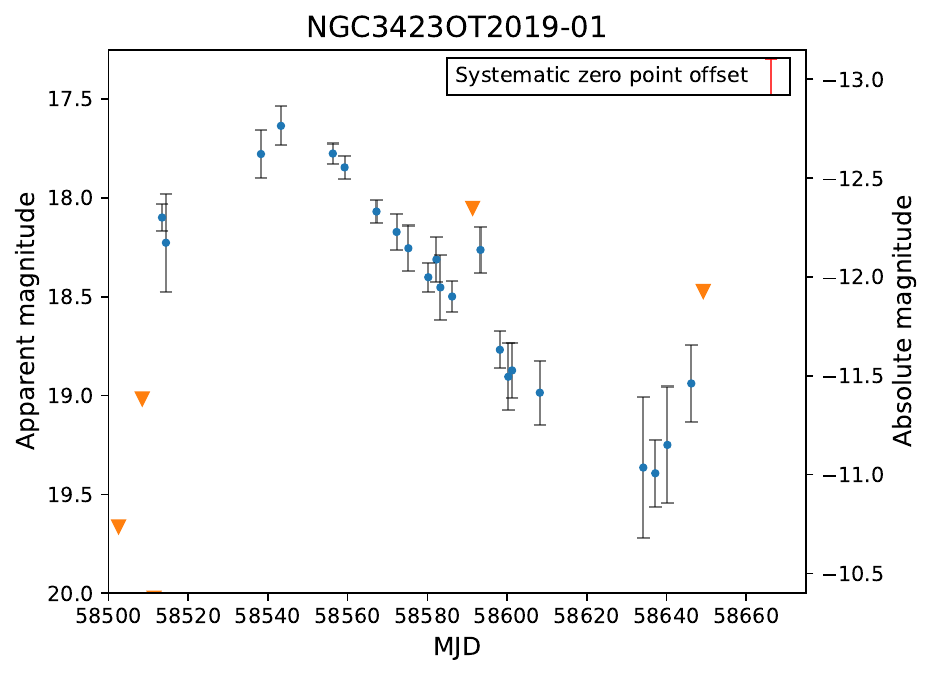}
    }
    \subfloat[]{
      \includegraphics[width=0.49\linewidth]{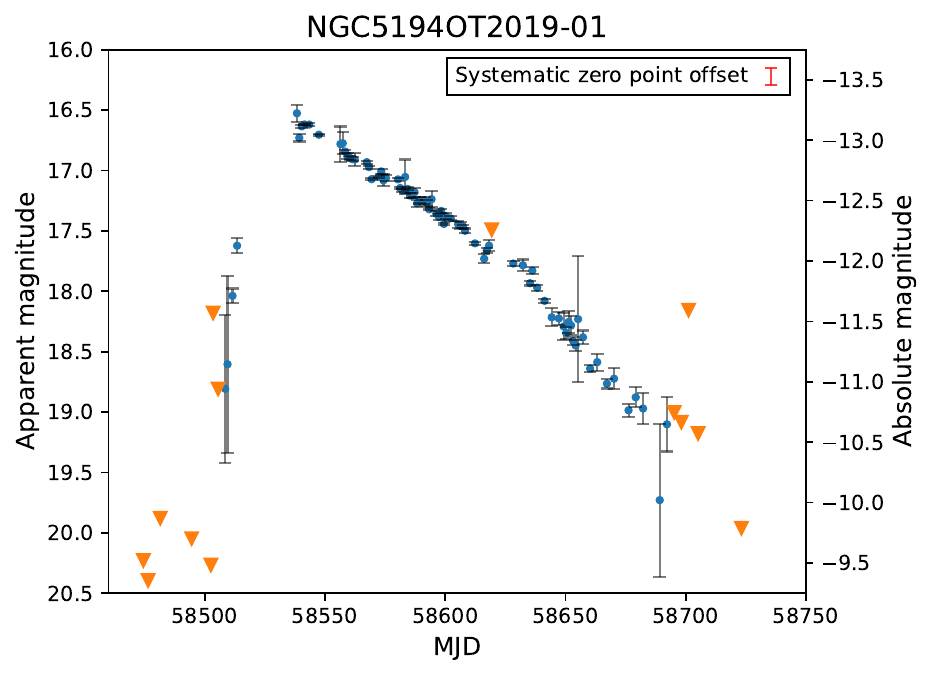}
    }
    
    \hspace{0mm}
    
    \subfloat[]{
      \includegraphics[width=0.49\linewidth]{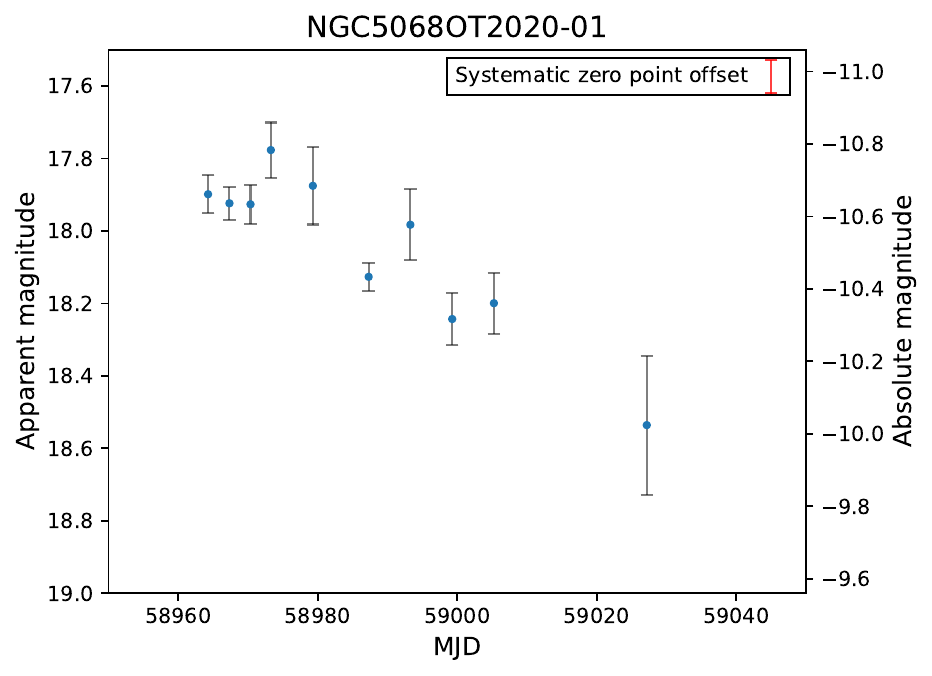}
    }
    \subfloat[]{
      \includegraphics[width=0.49\linewidth]{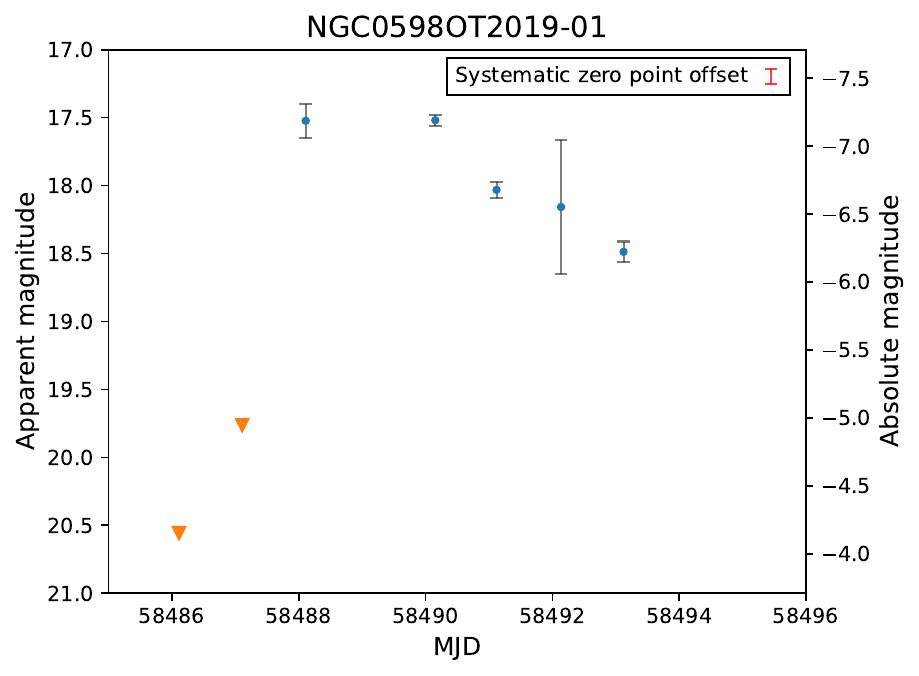}
    }
    \caption{Light curves of a sample of the identifiable sources found in the PTF/ZTF sample. (a): Type Ia supernova SN2020rcq. (b): Long-period variable. (c): Luminous Blue Variable AT2019ahd. (d): Intermediate Luminosity Red Transient AT2019abn. (e): Luminous Red Nova AT2020hat. (f): Nova AT2019gc. Photometric uncertainties on each observation are plotted. Orange triangles denote upper limits. All observations are normalised to the template; the uncertainty in the zero point for the template is illustrated as the potential offset of each light curve as a whole. Absolute magnitudes are approximate as kinematic distance moduli were taken from HyperLeda and extinction is not accounted for. A careful examination of distances and extinctions is used for unknown sources in Section \ref{subsec:unknown}}
    \label{fig:knownLCs}
\end{figure*}

The majority of sources recovered were either matched to previously known transients or recognisable as long-period variables. A sample of the light curves of these events is given in Figure \ref{fig:knownLCs}. 

\subsubsection{Supernovae}

The most common class of transient recovered was that of supernovae, accounting for six of the eleven sources found in the PTF data, and three of the eighteen sources found in the ZTF data. Six of these (SN2012cw, \citealp{2012cwtelegram}; SN2012fh, \citealp{2012fhtelegram}; SN2012P, \citealp{2012ptelegram}; iPTF13bvn, \citealp{13bvntelegram}; SN2018ivc, \citealp{2018ivctelegram}; SN2021gmj, \citealp{2021gmjtelegram}) are core-collapse supernovae, while the remaining three (SN2011ek, \citealp{2011ektelegram}; SN2013ct, \citealp{2013cttelegram}; SN2020rcq, \citealp{2020rcqtelegram}) are Type Ia events.

Additionally, one source recovered from ZTF observations of NGC 4725 is a background supernova candidate. This source is situated relatively far from the main body of the galaxy, implying that it may not be associated with it. Its coordinates do not match with any reported SNe. Due to its fast evolution, rising from below the detection threshold to its maximum luminosity in roughly 20 days before fading again from visibility in roughly 30 days, we suggest that this may correspond to a Type Ia supernova occurring in a dim background galaxy. 

\subsubsection{Long-period variables}

Four long-period variables were found, each in NGC 598 (M33). This was the nearest galaxy in our sample, allowing for the detection of dimmer events such as these. One variable was first discovered in PTF observations, without a corresponding detection in ZTF data. The remaining three were first discovered in ZTF observations, without corresponding detections from PTF. These three variables were dimmer, explaining their appearance in only the deeper ZTF images. In each case, photometry was carried out on the positions of each variable candidate using both the PTF and ZTF data.

ZTF analysis of the PTF source results in a detection, roughly 2 magnitudes dimmer, implying that this source may have reduced in brightness in the $\sim$ 5.6 year span between our final PTF detection and our ZTF detection. The ZTF sources are either missed entirely or only have marginal detections in the PTF data, presumably due to the shallower depth of these images.

\subsubsection{Asteroid}

One source recovered from PTF observations of NGC 1068 is an asteroid. This appeared as a very bright PSF-like source in a single image. Its measured position at time of observation matches with that of known asteroid 735 Marghanna\footnote{\url{https://minorplanetcenter.net/cgi-bin/checkmp.cgi}}. The asteroid was not visible at any position in the preceding or following images taken one day prior and four days later respectively. The pipeline requires a detection in two or more consecutive images to flag a source as potentially interesting. Upon checking, a spurious detection was found to have been made in the stack prior to the true detection of this asteroid. This source was found close to the centre of NGC 1068, where the quality of the subtraction was poorer, and so false detections were more likely to occur. Our single detection places this source at an apparent magnitude of 13.3.

\subsubsection{LBVs and supernova impostors}

Three sources recovered from ZTF observations correspond with known Luminous Blue Variables (LBVs), transients dimmer than supernovae arising from nonterminal outflows from massive stars. The LBVs detected are AT2019ahd in NGC 3423 \citep{at2019ahd}, AT2020pju in NGC 1385 \citep{at2020pju}, and AT2016blu in NGC 4559 \citep{at2016blu}. The analysis of AT2019ahd shows a rise and decline typical of such events. Only two observations of AT2020pju are obtained, both close the the detection threshold. AT2016blu is known to undergo frequent outbursts, our detections show a number of such events.

Additionally one source recovered from PTF observations of NGC 5806 corresponds with a previously known supernova impostor, SNHunt248 \citep{snhunt248}. Its evolution consists of two initial observations near the detection threshold, followed by two later brighter detections. This event has been modelled as a cool hypergiant undergoing a giant eruption similar to those of LBVs.

\subsubsection{ILRT}

One source recovered from ZTF observations of NGC 5194 (M51) corresponds with the Intermediate Luminosity Red Transient (ILRT) AT2019abn \citep{at2019abn}. These transients have been suggested to arise from the electron-capture induced explosion of a super-asymptotic giant branch star \citep{ILRTECSNE}.  Its light curve shows a rise to peak over roughly 30 days followed by a decline lasting roughly 150 days before passing below the detection threshold. Its decline consists of two linear segments, with a break occurring after 100 days.

\subsubsection{LRNe}

Two sources recovered from ZTF observations correspond with known Luminous Red Novae (LRNe), transients produced by the merger of a massive binary system \citep{LRNe}. The LRNe recovered are AT2018hso in NGC 3729 \citep{at2018hso} and AT2020hat in NGC 5068 \citep{at2020hat}. AT2018hso displays a quick rise to peak over 10 days and similar decline from visibility. The rise is not detected for AT2020hat, but its decline lasts longer, visibly decreasing in luminosity over 60 days before lack of imaging makes further observations impossible.

\subsubsection{Seyfert Galaxy}

One source recovered from ZTF observations corresponds with a known Type I Seyfert galaxy, 3XMM J140416.7+541615. This was detected during analysis of NGC 5457; its position on the sky was close enough to NGC 5457 to not be automatically rejected. It displays stochastic variability throughout observations. 

\subsubsection{Novae}

Two sources are found to correspond with previously known novae. PTF observations of NGC 598 rediscover known nova M33 2010-07a \citep{ngc0598nova}. This was the first nova observed in NGC 598 to undergo a second eruption. ZTF observations of NGC 3031 rediscover known nova M81 2018-11a \citep{ngc3031nova}.

\subsection{Unknown Sources}
\label{subsec:unknown}

A handful of candidates were found in the course of our search that could not be immediately identified as a known transient. We examined each of these candidates in detail, and in the course of this we reviewed the literature to determine the most reliable distance to their host. We list these distances and adopted foreground reddening in Table \ref{tab:host_extinction}.

\begin{table}
\centering
\begin{tabular}{@{}cccc@{}}
\toprule
Galaxy   & Distance Modulus & A$_\textrm{R}$ & A$_{\textrm{ZTF-r}}$ \\ \midrule
NGC 5194 & 29.67            & 0.075 & 0.079       \\
NGC 4258 & 29.39            & 0.035 & 0.037       \\
NGC 4826 & 28.34            & 0.090 & 0.095       \\ \bottomrule
\end{tabular}%
\caption{Adopted distances and foreground galactic extinction values for sources requiring further examination. NGC 5194 and NGC 4826 distances calculated using TRGB method \citep{ngc5194dist, ngc4826dist}. NGC 4258 distance calculated using maser method \citep{ngc4258dist}. Extinction values taken from NED. PTF observations are corrected using the value for the Landolt R filter. ZTF observations are corrected using the value for the SDSS r filter, taken as the closest match to the custom ZTF-r.}
\label{tab:host_extinction}
\end{table}

\subsubsection{NGC5194OT2020-01}

\begin{figure}
    \centering
    \includegraphics[width=\linewidth]{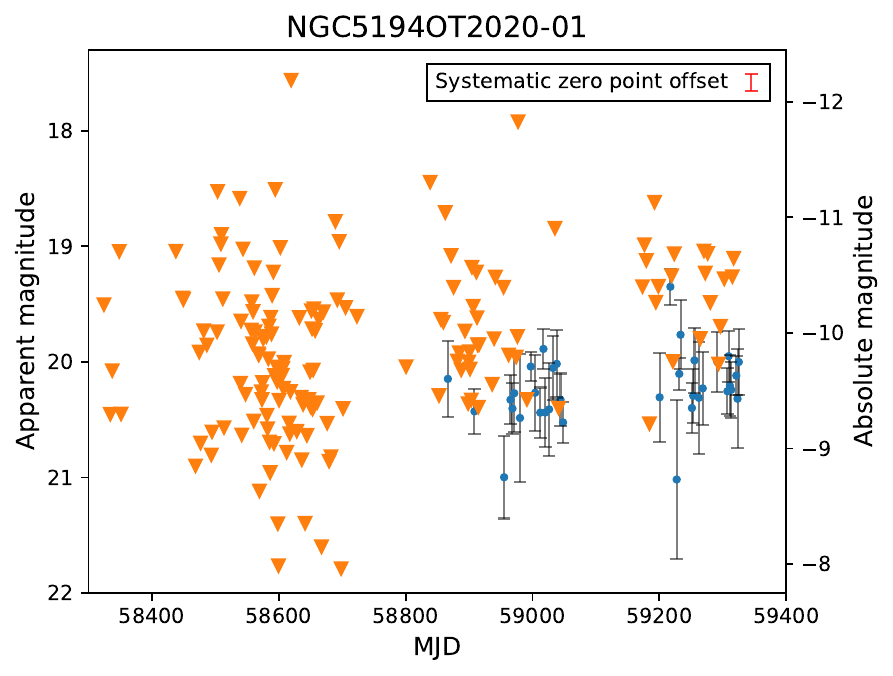}
    \caption{ZTF light curve for NGC5194OT2020-01. Orange triangles denote upper limits. Absolute magnitudes calculated using best estimate for distance and extinction correction in Table \ref{tab:host_extinction}.}
    \label{fig:ngc5194weirdoLC}
\end{figure}

\begin{figure}
    \centering
    \includegraphics[width=\linewidth]{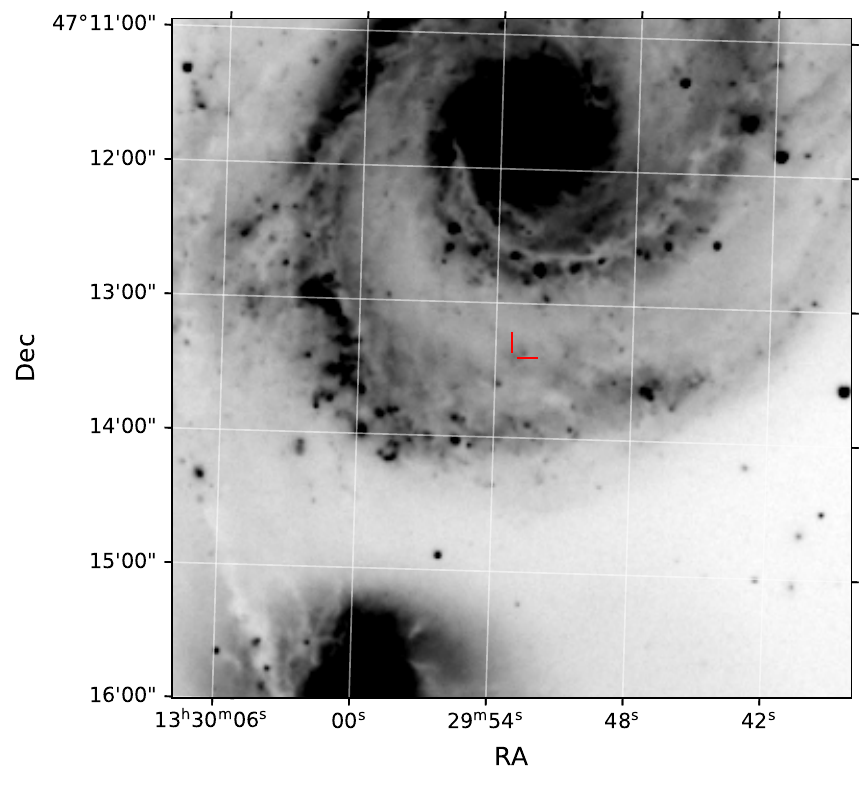}
    \caption{Position of NGC5194OT2020-01 in NGC 5194. Source is located 1.7 kpc from galaxy centre.}
    \label{fig:ngc5194weirdopos}
\end{figure}

\begin{figure}
    \centering
    \includegraphics[width=\linewidth]{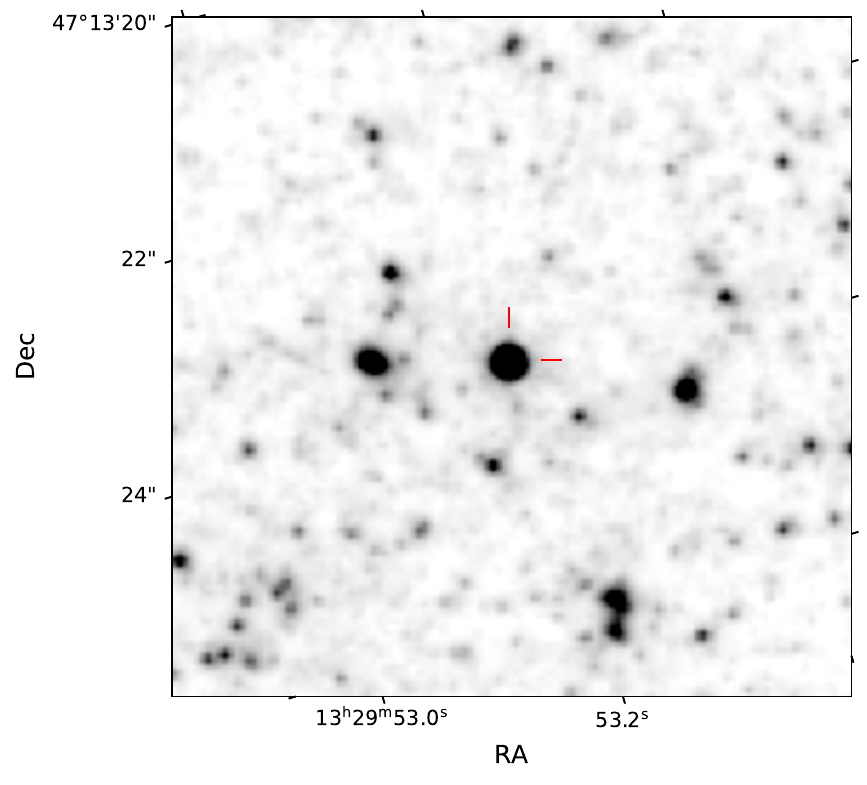}
    \caption{HST cutout of region around NGC5194OT2020-01.}
    \label{fig:ngc5194HSTpos}
\end{figure}

One source recovered from the ZTF photometry of NGC 5194 did not match any previously known transients. This warranted further analysis. The ZTF light curve of this source shows over a year of non-detections before the appearance of a source at MJD $\sim$ 58850 at an apparent magnitude of ZTF-r $\sim$ 20, corresponding to an absolute magnitude of $-9.7$. The source remains around this level of brightness, just above the detection threshold for the remainder of observation, corresponding to about 500 days. Our ZTF light curve of this event is presented in Figure \ref{fig:ngc5194weirdoLC}. The source is located roughly 2 arcminutes from the centre of the galaxy, within the main body between the two spiral arms. Its position in ZTF imagery is shown in Figure \ref{fig:ngc5194weirdopos}, while Figure \ref{fig:ngc5194HSTpos} shows a closer cutout on the position using {\it HST} images, with the source itself visible.

The intrinsic faintness of this transient, along with its maintenance of this magnitude for such a long period of time raised the possibility of this transient being associated with a failed supernova.

To further investigate this possibility, we searched for a potential progenitor to this event using {\it HST} data. Observations covering the position of this source were available for a number of epochs dating back to 2001. In each of the observations, a bright source was visible at the location returned from the pipeline, which we take to be the progenitor of the transient. We performed PSF-fitting photometry on each image using the {\sc DOLPHOT} code \citep{dolphot} with all parameters set to their recommended values from the {\sc DOLPHOT} handbook. Each set of images was analysed separately, using distinct deep drizzled images as reference frames. Examination of the residual images after PSF fitting and subtraction show no strong residuals, implying satisfactory fits in all cases. Magnitudes are corrected for extinction using values from NED.

A summary of the observations and the measured magnitudes of the source's progenitor are presented in Table \ref{tab:HST}. 

\begin{table}
\centering
\begin{tabular}{@{}ccccc@{}}
\toprule
Date        & Instrument    & Filter    & Exposure (s)  & Mag (err)         \\ \midrule
2001-06-09  & WFPC2         & F450W     & 4 x 500       & 21.615 (0.010)    \\
-           & -             & F555W     & 4 x 500       & 21.580 (0.008)    \\
-           & -             & F814W     & 4 x 500       & 21.331 (0.011)    \\
2005-01-19  & ACS/WFC       & F435W     & 1 x 680       & 20.983 (0.005)    \\
-           & -             & F555W     & 1 x 340       & 20.808 (0.007)    \\
-           & -             & F814W     & 1 x 340       & 20.481 (0.007)    \\
2005-01-20  & ACS/WFC       & F435W     & 1 x 680       & 20.978 (0.005)    \\
-           & -             & F555W     & 1 x 340       & 20.800 (0.007)    \\
-           & -             & F814W     & 1 x 340       & 20.515 (0.007)    \\
2005-01-21  & ACS/WFC       & F435W     & 1 x 680       & 20.984 (0.005)    \\
-           & -             & F555W     & 1 x 340       & 20.807 (0.007)    \\
-           & -             & F814W     & 1 x 340       & 20.508 (0.007)    \\
2005-01-22  & ACS/WFC       & F435W     & 1 x 680       & 20.983 (0.005)    \\
-           & -             & F555W     & 1 x 340       & 20.809 (0.007)    \\
-           & -             & F814W     & 1 x 340       & 20.500 (0.007)    \\
2019-05-31  & ACS/WFC       & F814W     & 4 x 564       & 21.240 (0.004)    \\
2021-04-28  & ACS/WFC       & F606W     & 4 x 552       & 20.061 (0.001)    \\
2021-04-29  & ACS/WFC       & F814W     & 4 x 552       & 19.760 (0.002)    \\ \bottomrule
\end{tabular}
\caption{{\it HST} photometry for progenitor of NGC5194OT2020-01.}
\label{tab:HST}
\end{table}

\begin{figure}
    \centering
    \includegraphics[width=\linewidth]{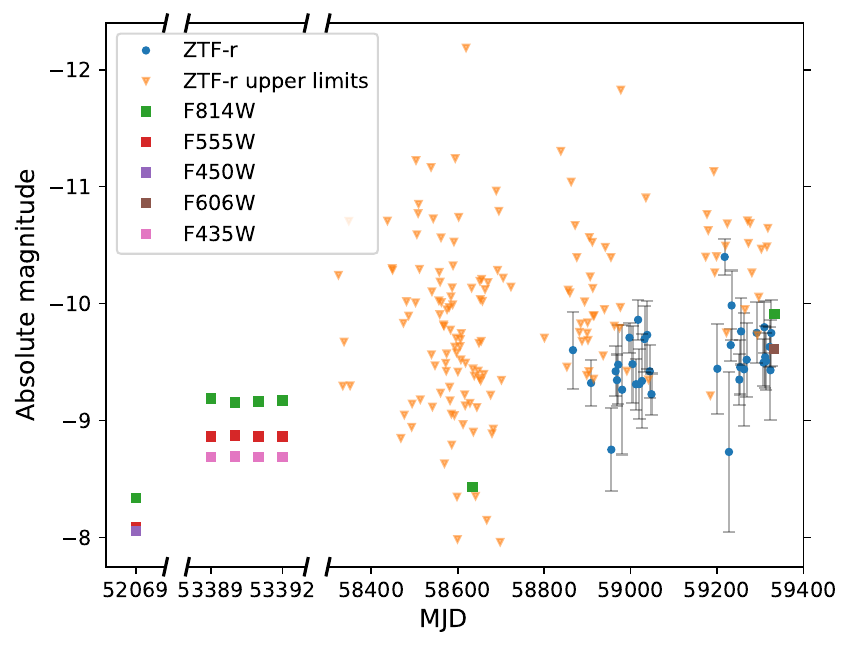}
    \caption{Combined ZTF and {\it HST} light curve for NGC5194OT2020-01. Squares mark observations from {\it HST}, circles mark ZTF observations, and triangles indicate upper limits on ZTF observations. X-axis includes two breaks, with the scale differing in each segment of the light curve. First segment shows first available detection from {\it HST} in 2001. Second segment shows four consecutive days of observation from {\it HST} in 2005. Final segment comprises entirety of ZTF observations, as well as {\it HST} detections from both 2019 and 2021, amounting to $\sim$ 1000 days. Absolute magnitudes calculated using best estimate for distance and extinction correction in Table \ref{tab:host_extinction}. {\it HST} observations corrected for extinction using values from NED.}
    \label{fig:ngc5194weirdofullLC}
\end{figure}

Figure \ref{fig:ngc5194weirdofullLC} shows the long term evolution of this source. Each epoch of observation includes measurements in $F814W$, so we use this to track the overall photometric evolution. In our first observations in 2001, the source shows a magnitude of 21.331 $\pm$ 0.011. By the time of the next set of observations, the source has brightened by almost a magnitude to 20.500 $\pm$ 0.007 on 2005-01-22. Following this, no observations are available until 2019, by which point the source has dimmed to a magnitude of 21.240 $\pm$ 0.004, essentially matching its 2001 level. Finally, by 2021, the source brightens by $\sim$ 1.5 magnitude to reach 19.760 $\pm$ 0.002, 0.74 magnitudes above its level in 2005. Taking a distance modulus of 29.16 from NED as in Table \ref{tab:host_extinction}, this corresponds to a peak absolute magnitude of -9.4 in the F814W band.

We refer to the source's state in 2001 as its quiescent phase, with the observations in 2005 and 2021 occurring during periods where the source is brighter than usual. As the F814W observations from 2019 match those of 2001, we assume the source to be quiescent at this point too.

We calculate the change in the $F814W - F555W$ colour between the 2001 and 2005 epochs. This is the only instance of photometry being available in multiple bands at multiple epochs. For this, we correct our observed magnitudes of reddening using values taken from NED. In 2001, the candidate has an $F814W - F555W$ colour of $-0.205\pm0.014$ mag, while in 2005 when it is brighter this colour is $-0.265\pm0.01$ mag. In both epochs, the source is red, and it is redder in the 2005 observation. If the variability in brightness between these two epochs was due to dust, we would expect that the source would be redder in 2001, when it is fainter. This disfavours variability in dust as the cause of the 2005 brightening. 

We use Hoki \citep{hoki}, a Python module designed to interface with BPASS models \citep{bpass}, in order to determine the parameters of the system which best reproduces our observations. We focus on matching the observations taken in 2001, these representing a quiescent point in the system's evolution during which we have observations in multiple bands. We consider both single-star and binary systems at solar metallicities. In each case, we find the BPASS models which match the $F450W$, $F555W$, and $F814W$ magnitudes from the 2001-06-09 observations the closest. Again, we correct our magnitudes using the extinctions from NED and convert to absolute magnitudes using the best estimate for the distance modulus from the same source.

For both the single-star and the binary case, we find no Hoki models which match our observations to within their photometric uncertainties. This may occur due to the combination of the low photometric uncertainties offered by the {\it HST} observations and the discreteness of the BPASS models being used. We arbitrarily increase our leniency to search for all models matching our extinction-corrected observations to within 0.05 mag in each band.

\begin{figure}
    \centering
    \includegraphics[width=\linewidth]{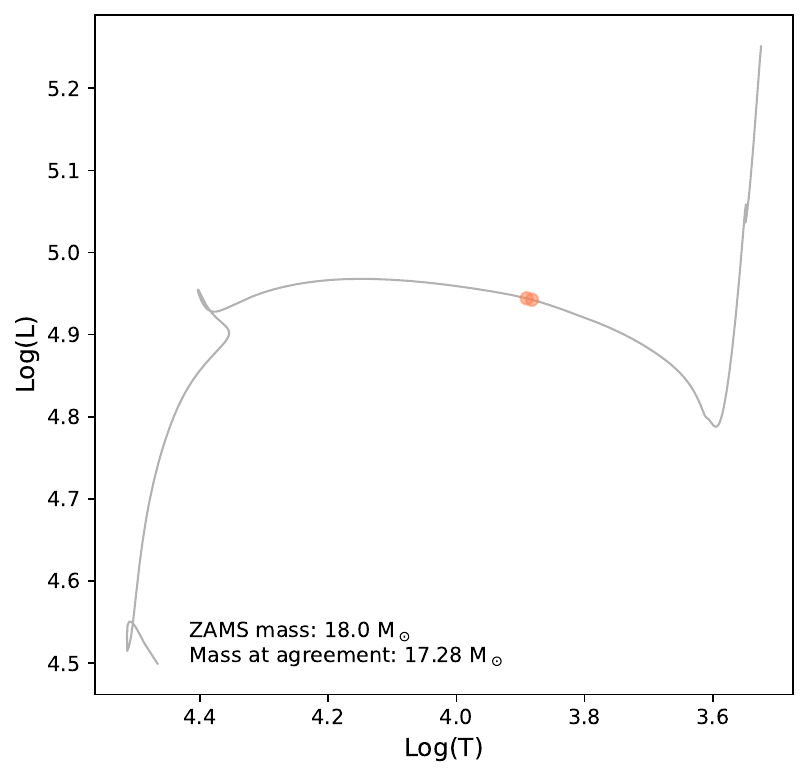}
    \caption{Evolutionary track of best fitting single-star Hoki model to NGC5194OT2020-01. Orange spots refer to the points in evolution where observations in $F450W$, $F555W$, and $F814W$ bands agree with model to within 0.05 mag.}
    \label{fig:ngc5194single}
\end{figure}

For the single-star case, we find a single model which matches our observations in each band to within 0.05 magnitudes. This comes from a star with a Zero Age Main Sequence mass of 18 M$_\odot$. At the point in its evolution where it agrees with our observations it has a mass of 17.28 M$_\odot$ with $\log L/L_\odot = 4.9$~dex and $\log T = 3.9$~dex. It is evolving redward, and is expected to end its life as a red supergiant. The evolutionary path of this model is shown in Figure \ref{fig:ngc5194single}, with the positions at which the model agrees with observations highlighted.

\begin{figure}
    \centering
    \includegraphics[width=\linewidth]{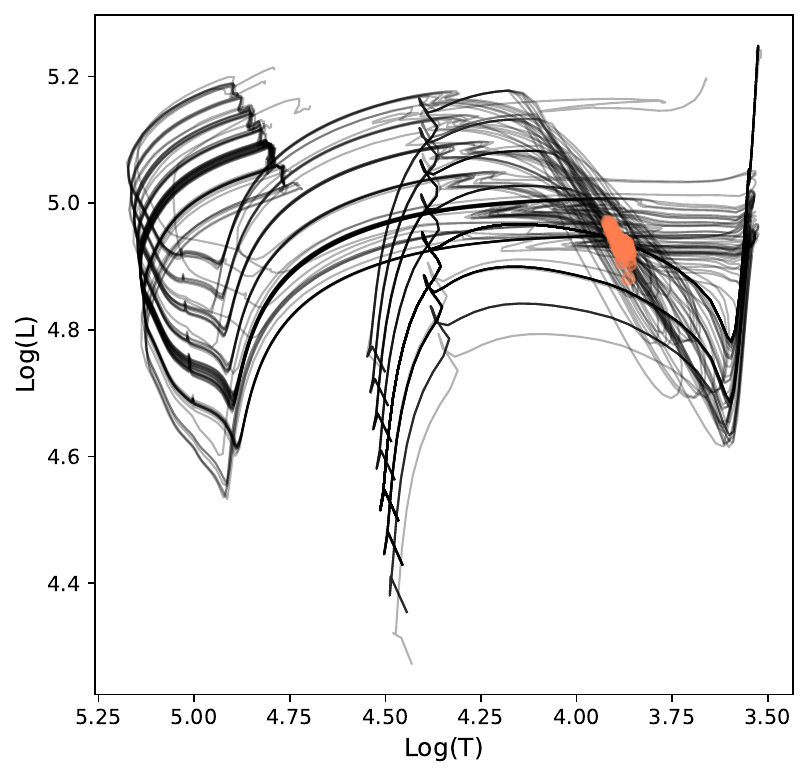}
    \caption{Evolutionary tracks of 96 best fitting binary Hoki models to NGC5194OT2020-01. Orange spots refer to the points in evolution where observations in $F450W$, $F555W$, and $F814W$ bands agree with any model. Darker lines show locations where larger number of models agree with each other. Mass of primary component ranges from 15 - 22 M$_\odot$.}
    \label{fig:ngc5194binary}
\end{figure}

For the binary case, the increase in the number of free parameters allows for more models to agree with our observations. We find 96 models with luminosities matching our observations to within 0.05 magnitudes in each band. Each model's evolution is shown in Figure \ref{fig:ngc5194binary}, with the positions at which the models agree with observations highlighted. 

Some fraction of massive stars experience an LBV-like phase in their evolution. The physics governing this is currently poorly understood, but stars in this phase can undergo variability potentially consistent with our observations. This points towards this source perhaps being similar to the yellow hypergiant IRC+10420. Particularly, the temperature of our single-star model of $\sim8000$~K is close to the temperature of $\sim$9200 K reported by \cite{irc10420}.

\subsubsection{NGC4258OT2010-01}

\begin{figure}
    \centering
    \includegraphics[width=\linewidth]{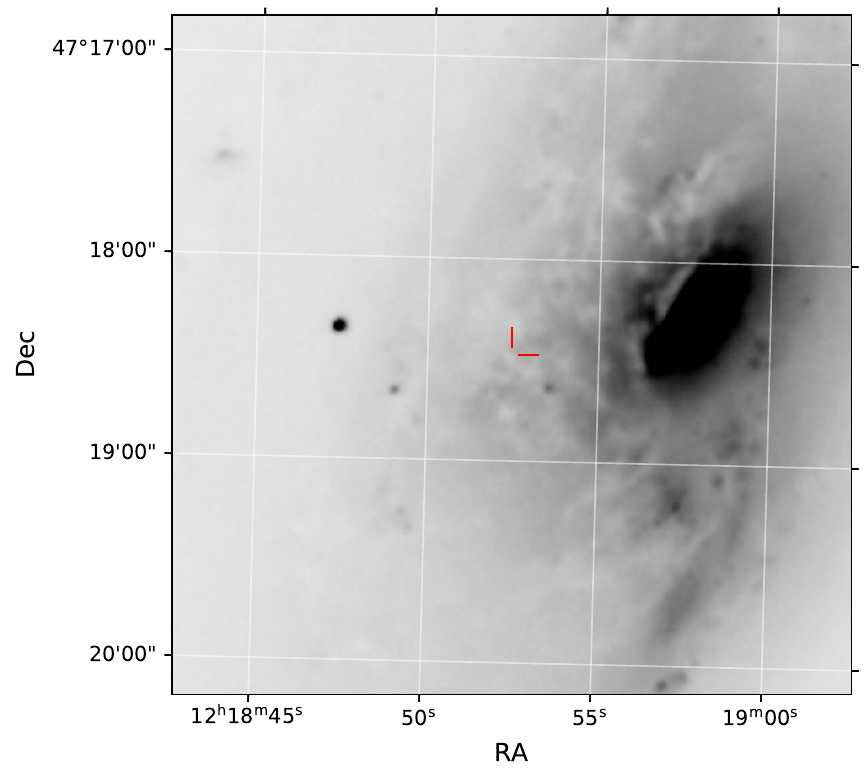}
    \caption{Position of NGC4258OT2010-01 in NGC 4258.}
    \label{fig:ngc4258weirdopos}
\end{figure}

\begin{figure}
    \centering
    \includegraphics[width=\linewidth]{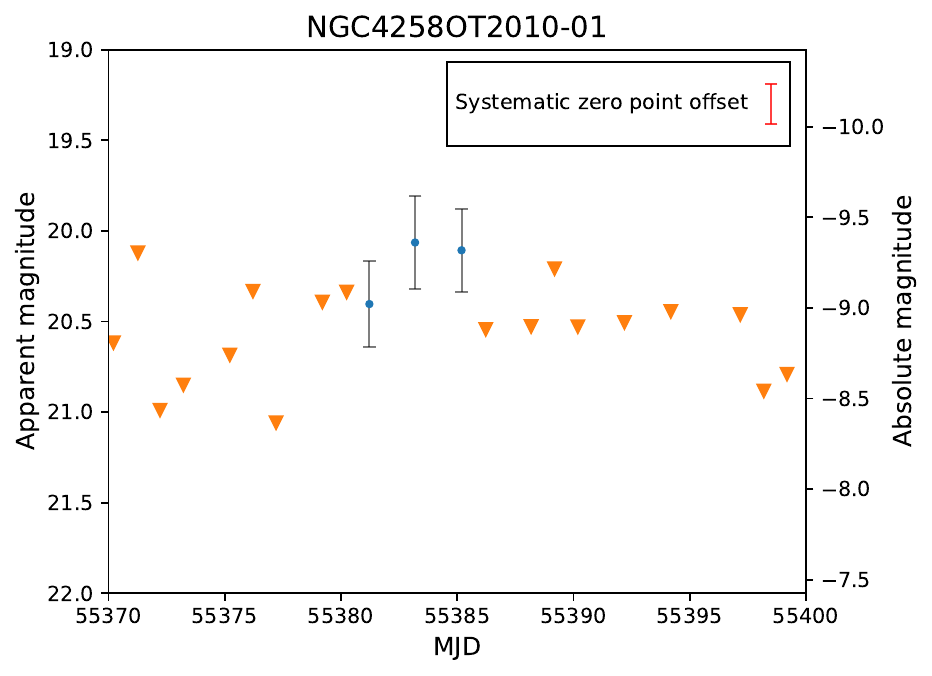}
    \caption{PTF light curve for NGC4258OT2010-01 produced using AutoPhoT code. Absolute magnitudes calculated using best estimate for distance and extinction correction in Table \ref{tab:host_extinction}.}
    \label{fig:ngc4258weirdoLC}
\end{figure}

Further analysis was performed on one source from PTF photometry of NGC 4258 which matched no previously known transients or variables. The position of this source in its host galaxy is shown in Figure \ref{fig:ngc4258weirdopos}. The PTF light curve for this event consists of three faint detections. Our first detection is at magnitude 20.4, with the next two appearing at magnitude 19.8. Each of these detections are separated by two days. Our original photometry showed a limiting magnitude of $\sim22$ two days prior to first detection which would imply an extremely fast rise in magnitude over a very short time. To check this, further photometry was carried out using AutoPhoT \citep{autophot}. The light curve produced from the AutoPhoT photometry is shown in Figure \ref{fig:ngc4258weirdoLC}.

\begin{figure}
    \centering
    \includegraphics[width = \linewidth]{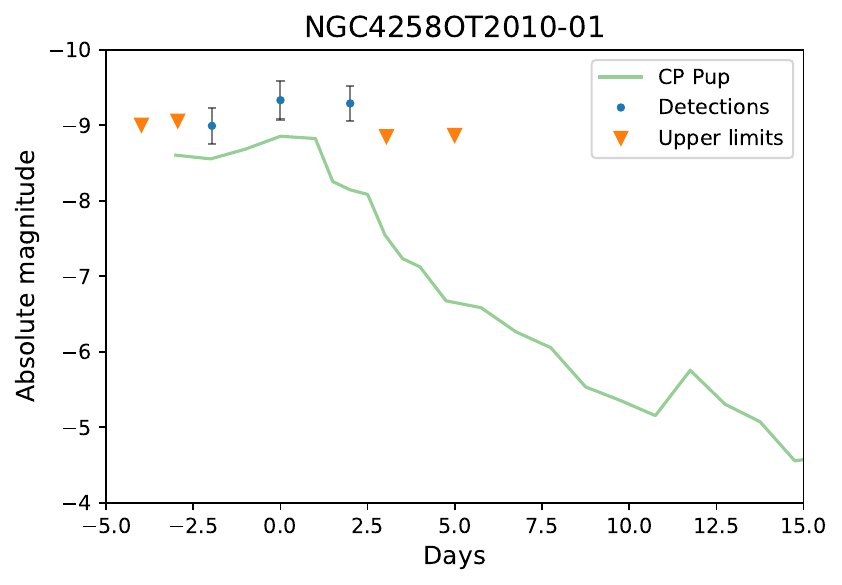}
    \caption{Comparison of NGC4258OT2010-01 to nova CP Pup. Brightness and shape of this nova are plausibly consistent with our transient.}
    \label{fig:ngc4258nova}
\end{figure}

AutoPhoT returns limiting magnitudes prior to the first detection much closer to the detection magnitude. This refutes the constraint that a very fast rise time is required to explain this transient. Due to its short duration and relatively low absolute magnitude, we suggest that this event may correspond with the peak of a nova. We compare the shape of the three detections along with the limiting non-detections before and after with the shapes of light curves of 93 novae from \cite{novaLCs}. For a number of novae whose shapes match our candidate most closely, we take distances from Gaia parallax measurements \citep{gaianovae} to convert to absolute magnitudes. We align the nova light curves such that their maximum brightness occurs at the time of the candidate's second detection. Figure \ref{fig:ngc4258nova} shows a comparison of our light curve with that of nova CP Pup.

CP Pup shows a shape and magnitude consistent with our transient. While this is based on only three detections and relatively non-constraining limiting magnitudes, the similarities are enough for us to classify this event as a potential nova.

\subsubsection{NGC4826OT2019-01}

\begin{figure}
    \centering
    \includegraphics[width=\linewidth]{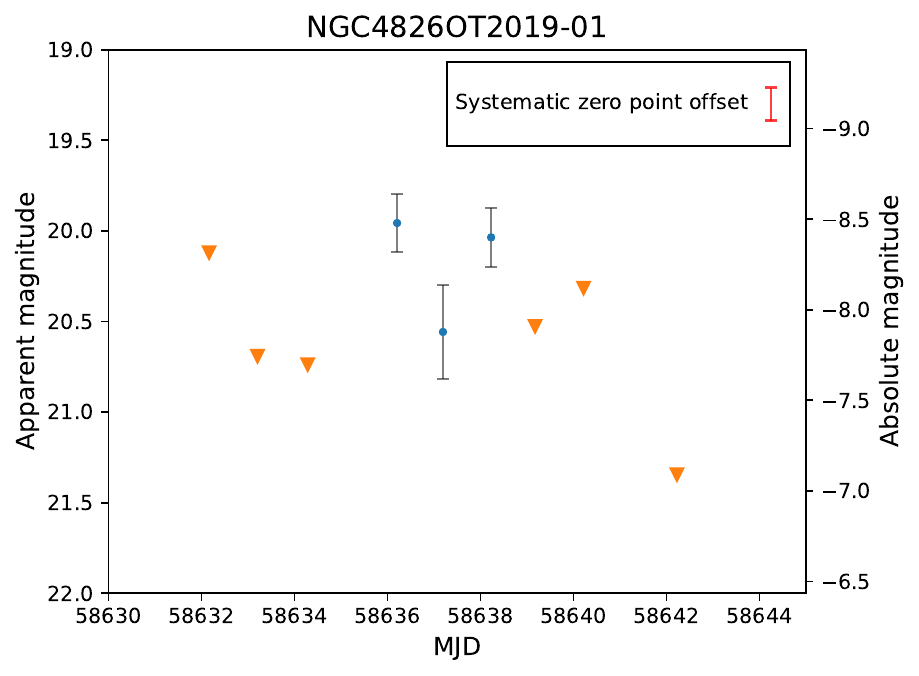}
    \caption{Light curve for NGC4826OT2019-01. Absolute magnitudes calculated using best estimate for distance and extinction correction in Table \ref{tab:host_extinction}.}
    \label{fig:ngc4826weirdoLC}
\end{figure}

Further analysis was performed on one source from ZTF photometry of NGC 4826 which matched no previously known transients or variables. The ZTF light curve for this event consists of three faint detections around or below magnitude 20, each separated by a day. Magnitudes one day prior to first detection and one day post final detection are limited to magnitudes 20.7 and 20.5 respectively. The ZTF light curve is presented in Figure \ref{fig:ngc4826weirdoLC}.

\begin{figure}
    \centering
    \includegraphics[width=\linewidth]{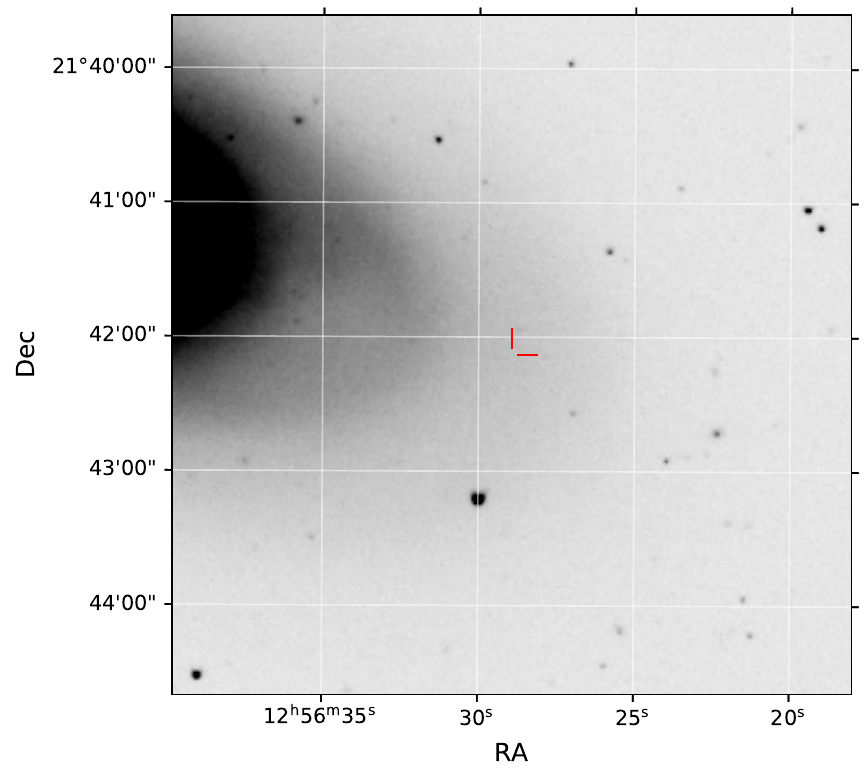}
    \caption{Position of NGC4826OT2019-01 in NGC 4826. The transient is far from any obvious regions of active star formation.}
    \label{fig:ngc4826weirdopos}
\end{figure}

The source is located near the edge of its host galaxy, as shown in Figure \ref{fig:ngc4826weirdopos}, indicating a position within a region experiencing low levels of star formation. This would indicate a low probability of this event arising from a young progenitor, thus it is unlikely that this transient corresponds with a failed supernova event from a red supergiant.

\begin{figure}
    \centering
    \includegraphics[width=\linewidth]{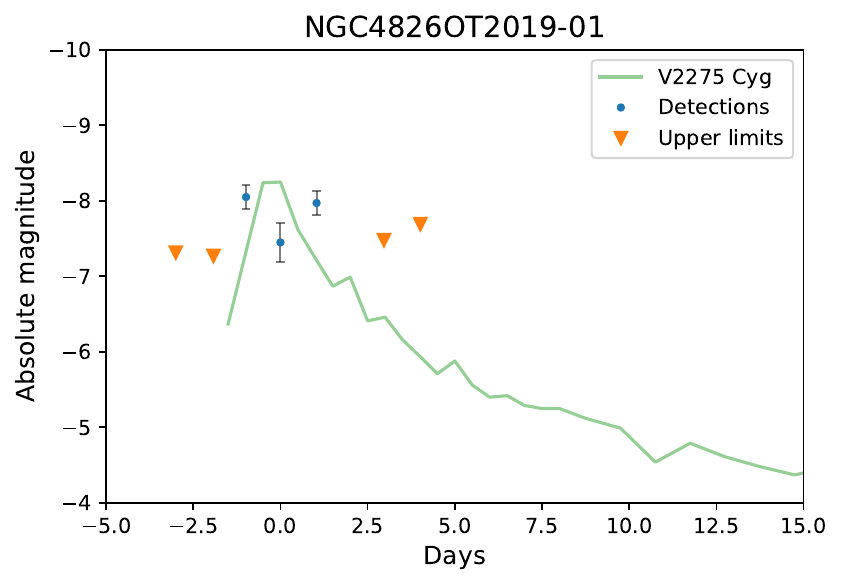}
    \caption{Comparison of NGC4826OT2019-01 to nova V2275 Cyg. Brightness and shape are plausibly consistent with our transient.}
    \label{fig:ngc4826nova}
\end{figure}

Similar to NGC4258OT2010-01, this object is not constrained to a particularly fast rise or decline, and appears for only a short time at a relatively dim absolute magnitude. We compare this transient with the same set of novae, and show a comparison to V2275 Cyg in Figure \ref{fig:ngc4826nova}.

The overall shape and magnitude of this source matches V2775 Cyg reasonably well. As with our previous source, we have relatively few detections but suggest this source be classified as a potential nova.


\section{Rates of failed SNe} \label{sec:rates}

Our re-analysis of PTF/ZTF observations for nearby galaxies found no promising failed supernova candidates. Using this non-detection, we place a constraint on the upper limit of the rate of failed supernovae, which we can then compare to the rate of core-collapse supernovae in the same sample.

The overall B-band luminosity of each galaxy was calculated. B-band magnitudes were taken from HyperLeda, and were converted to values in terms of solar luminosity, taking 5.44 as the sun's absolute B-band magnitude \citep{solarmag}. For both surveys, each luminosity was multiplied by the fraction of the galaxy in question visible in its master template to account for the loss in observed luminosity for galaxies only partially within the field of view. Each of these values were then multiplied by the number of years of data available for each galaxy in each survey. Finally, these values were multiplied by the calculated recovery efficiency for failed supernovae of each magnitude for the galaxy in question. This gave us an effective detection rate for both PTF and ZTF observations in units of $B$-band solar luminosity-years per galaxy, accounting for the efficiency of the pipeline in recovering failed supernovae.

We assume that failed supernovae occur as a Poissonian process at an intrinsic rate of $\lambda$, with a probability P of observing k events, as given by Equation \ref{eq:Poisson}.

\begin{equation}
    \label{eq:Poisson}
    P = \frac{e^{-\lambda}\lambda^k}{k!}
\end{equation}

There is a 5 per cent chance of finding 0 failed supernovae in the sample if the intrinsic rate $\lambda$ is 3 over the entire survey. To convert this to a 95 per cent confidence upper bound on the rate of failed supernovae in terms of B-band solar luminosity-years covered, we take the inverse of the sum of B-band solar luminosity-year measurements for each galaxy, and multiply this result by 3. The calculated rates for failed supernovae of each tested magnitude per $10^9$ B-band solar luminosity-years are given in Table \ref{tab:rates}.

We compare these rates to the core-collapse supernova rate seen within the same data set. This rate is calculated from the number of CCSNe discovered in our sample throughout the time covered. A total of 16 CCSNe occur in galaxies present in our sample within the time ranges of the PTF and ZTF surveys. Interestingly, 14 of these occur during PTF, while only 2 occur during ZTF. These include a number of supernovae which take place at points where PTF imaging was sparse, causing them to be missed. The parameter space for these events is taken as the entirety of the B-band solar luminosity-years observed across all galaxies for both surveys. The calculated upper limits on the ratio of the failed supernova rate compared to the CCSN rate are given in Table \ref{tab:rates}.

For example, if failed supernovae are expected to produce a transient with an absolute magnitude of $-11$, we can say with 95 per cent confidence that the rate of failed SNe must be $\leq$ 0.609 times that of core-collapse supernovae.

\begin{table*}
    \centering
    \resizebox{\textwidth}{!}{%
    \begin{tabular}{@{}cccccc@{}}
        \toprule
        Failed SN Magnitude & PTF Recovery Efficiency & ZTF Recovery Efficiency & Total $10^9$ L$_\odot$-years observed & 95\% Upper Bound on  & 95\% Upper Bound on Ratio of \\
                            &                         &                         &                                       & Failed SN Rate       & Failed SN to CCSN Rate \\ \midrule
        -11 & 0.246 & 0.331 & 4454  & 6.74 $\times 10^{-4}$ & 0.609 \\
        -12 & 0.501 & 0.636 & 8297  & 3.62 $\times 10^{-4}$ & 0.327 \\
        -13 & 0.639 & 0.784 & 10329 & 2.90 $\times 10^{-4}$ & 0.262 \\
        -14 & 0.710 & 0.865 & 11624 & 2.58 $\times 10^{-4}$ & 0.233 \\ \bottomrule
    \end{tabular}%
    }
    \caption{95 per cent upper bound on rate of failed supernovae per $10^9$ B-band solar luminosity-years for different magnitudes, and ratio of rate of failed supernovae to that of core-collapse supernovae.}
    \label{tab:rates}
\end{table*}

\begin{figure}
    \centering
    \includegraphics[width=\linewidth]{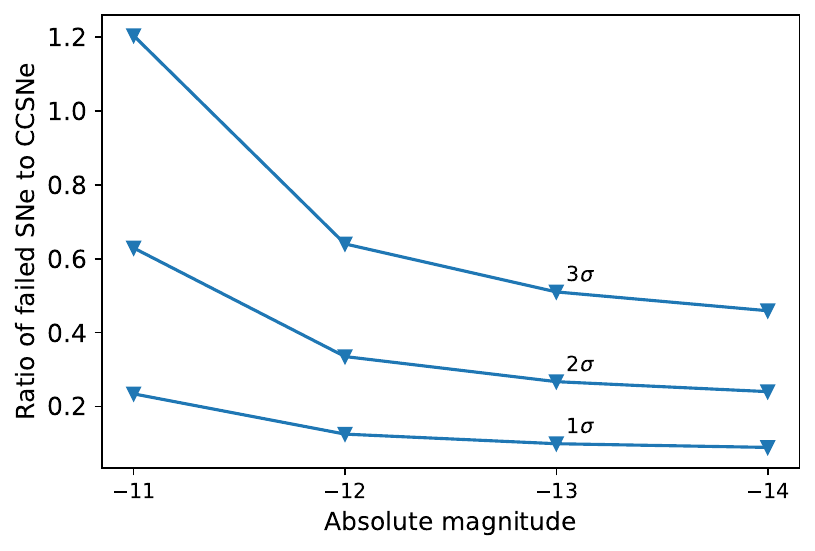}
    \caption{Exclusion plot illustrating upper limits on ratio of rate of failed supernovae of different magnitudes to that of core-collapse supernovae at different significance levels. At the significance levels indicated by the text above each contour, we expect the true ratio of failed supernovae to core-collapse supernovae to lie below the contour.}
    \label{fig:exclusion}
\end{figure}

Using the same method as above, we calculate upper limits on the ratio of failed supernovae to CCSNe at $1\sigma$, $2\sigma$, and $3\sigma$ confidence levels at each magnitude. These are displayed as an exclusion plot in Figure \ref{fig:exclusion}.

As Type II supernovae are expected to arise from progenitors with masses greater than 8 M$_\odot$, we can examine the Salpeter initial mass function \citep{salpeter} to gain a rough estimate of the proportion of failed supernovae to Type II supernovae. Integrating the IMF from $8-16$ M$_\odot$, corresponding to the range of Type II progenitors we see, and comparing this to the integral evaluated between $16-30$ M$_\odot$, we find that roughly one quarter of RSGs will have masses above this threshold. Under the naive assumption that all RSGs less massive than this will end their lives as CCSNe and all RSGs above this will produce failed SNe, the expected ratio of failed to Type II supernovae would thus be 0.33. As there will also be some massive stars that produce Type Ibc SNe, the actual ratio compared to all CCSNe will be lower than this. While there is no hard cut-off point, and factors other than mass play a large role in determining the explodability of a star \citep{pattonsukhbold}, this is a useful figure to consider as a first approximation. Our calculated ratios lie below this value for failed SNe of magnitude -12 or brighter. It is therefore likely that we would have observed a failed supernova in our sample if they did occur at this magnitude. For our non-detection to be consistent with this assumed ratio, we propose that failed supernovae must be produced at absolute magnitudes equal to or dimmer than -11.

We also note that \citep{neustadt2021} recently reported a failed SN fraction based on their search for disappearing massive stars of $0.16^{+0.23}_{-0.12}$ at 90 per cent confidence, which is consistent with what find here.

\section{Discussion} \label{sec:discussion}

Our analysis suggests that it is unlikely that failed supernovae could produce a transient of absolute magnitude $-12$ or brighter in the $R$ or ZTF-$r$ bands. However, our non-detections cannot rule out failed supernovae of absolute magnitude $-11$ or lower. From the expectation that the failed supernova rate is less than one third that of Type II SNe, and the knowledge that Type II SNe account for roughly 70 per cent of the total population of CCSNe \citep{typeIIrate}, we can deduce that the rate of failed supernovae should be less than 0.23 times the total CCSN rate. Our efficiency calculations show that we expect to recover 25 to 33 per cent of failed supernovae at magnitude $-11$. With 16 CCSNe in the host sample, this results in an expected number of magnitude $-11$ failed supernova detections between 0.9 and 1.2. At such a low expected number, detecting zero is not necessarily surprising. For any potential failed supernovae brighter than this, the expected numbers are higher, leading us to conclude that failed supernovae do not produce transients of absolute magnitude $-12$ or higher. This brightness cut-off is not necessarily surprising.

The \citeauthor{lovegrove2013} model which is use as our template failed supernova light curve peaks at a bolometric luminosity of $\sim 6 \times 10^{39} \textrm{ erg s}^{-1}$. We use PySynphot \citep{pysynphot}, a Python package for simulating observed spectra and light curves, to investigate the expected luminosities of such a transient as observed by the Vera Rubin Observatory's Legacy Survey of Space and Time \citep[LSST;][]{lsst}. We generate three blackbody spectra at temperatures of 3000 K, 4000 K, and 5000 K, each scaled to a luminosity of $6 \times 10^{39} \textrm{ erg s}^{-1}$. We simulate observations of each of these spectra through the {\it ugrizy} filters of LSST. We find that each of these sources would peak in the {\it y} band. Those with temperatures of 3000 K would peak in this band at an absolute magnitude of -11.42. Sources with temperatures of 4000 K and 5000 K would peak at -11.42 and -11.24 respectively.

In addition, our results rely on the shape of the light curve presented by \citeauthor{lovegrove2013}. All efficiency calculations essentially measure the likelihood of our pipeline recovering a transient of this shape lasting 300 days. In reality, it is likely that failed supernova events may occur on timescales shorter or longer than this, which would affect their detectability. For example, a longer lasting failed supernova may be less likely to be missed between observing cycles. However, an extension of its light curve may cause its peak luminosity to drop below the detection threshold, causing it to be missed entirely. Many factors relating to the length and shape of the light curve would have important effects on the detectability of these events.

We expect these events to originate from very massive red supergiants, accounting for the missing high-mass red supergiant progenitors of Type IIP supernovae. However, it is possible that these massive red supergiants evolve further to become yellow supergiants prior to core collapse. The dim optical transient from a failed supernova is expected to come from the recombination of hydrogen as the outer layers of the star are ejected post-collapse of the core to a black hole. This is possible for red supergiants due to their tenuously bound extended outer layers, but yellow supergiants are much more tightly bound. As such, the production of an optical transient would be much more difficult. This path may explain the lack of Type IIP supernovae from high-mass progenitors and the difficulty of detecting optical transients from failed supernovae.

An additional complicating factor is the impact of binaries. Our models assume these transients are produced from single stars. In reality, more than 70 per cent of massive stars may exist in binaries and undergo interaction with a companion during their lifetime \citep{binaries}. Such interactions can result in the stripping of a star's outer layer, leading to the production of supernovae which look markedly different from those produced by single stars. It is likely that similar processes could affect stars which would produce failed supernovae similar to those of our models if isolated, resulting in the production of transients which may be even dimmer, or evolve differently. Additionally, binary stripping can result in a smaller core mass and compactness, which would enhance the explodability of a star \citep{laplace}. Accounting for these effects, detectable failed supernovae from red supergiants may be even less common.

The future detection of optical transients from failed supernovae or further constraints on their rates will be helped immensely by the availability of further deep, high cadence imagery of nearby galaxies. In the near future, the LSST at the Vera Rubin Observatory will commence its 10 year survey of the sky at a depth of r $\sim$ 24.5. This observatory will allow us to detect even dimmer transients, and to greater distances than before.

We calculate the distance to which a failed supernova would be detectable from the Vera Rubin observatory. To do this, we begin with the typical $5\sigma$ source depths for LSST \citep{lsst}. To select only for events which would be easily distinguishable as a failed supernova, as require that failed SNe must peak two magnitudes above the limiting magnitude of the survey. We then compare these conservative depths with the peak magnitudes predicted by our PySynphot models in each band.

A failed supernova with a bolometric luminosity of $6 \times 10^{39} \textrm{ erg s}^{-1}$ at a temperature of 3000 K would be comfortably detectable in LSST's {\it i} filter at a distance modulus of 32.6, corresponding to a distance of 33 Mpc. For source at 4000 K, this detectability extends to 39 Mpc in the {\it i} band, while a 5000 K source would be detectable to a distance of 43 Mpc in the {\it r} band.

\cite{typeIIrate} calculate a local volumetric rate of Type II SNe of $0.447 \times 10^{-4} \textrm{ MPc}^{-3} \textrm{ year}^{-1}$. This implies that an average of 6.6 Type II SNe should occur within 33 Mpc per year. At the Cerro Pachón site of the Vera Rubin observatory, roughly 75 per cent of the sky will be visible, meaning that 5 of these would be visible to LSST. Assuming that failed supernovae occur at one third the rate of Type II SNe (as discussed in Section \ref{sec:rates}), this would suggest that an average of 1.7 failed supernovae would be visible within LSST every year, assuming the transient occurs at 3000 K. This would translate to 17 such events over the entirety of the 10 year survey. For failed supernovae occurring at 4000 K or 5000 K, a total of 2.8 or 3.7 events would be visible on average each year.

Given our initial assumptions on the ratio of failed supernovae to Type II SNe, our derived rates are unfortunately not very constraining. From the slope of the IMF and the observed population of Type IIP supernova progenitors, we crudely expect a ratio of failed supernovae to CCSNe of $\sim$0.23. As such, our constraints (i.e. that the upper limit to the relative rate of failed supernovae is between 0.61 and 0.23 from absolute magnitudes of -11 and -14 respectively) are not very restrictive. In this light, the lack of a detected failed supernova is perhaps not surprising. Nonetheless, these limits are to our knowledge the only direct constraints on the rate of Lovegrove and Woosley-like transients. We suggest that applying this methodology and search strategy to larger data sets from future surveys will provide a promising avenue.

\section{Conclusions} \label{sec:conclusions}

We present results from a new analysis pipeline designed to search for faint transients associated with failed supernovae in the combined PTF/ZTF data set. Our analysis rediscovered previously known transients, and found a small number of previously unknown candidates. Two of these events are consistent with novae, while the third displays a long-term outburst potentially indicative of an LBV. No promising failed supernova candidates are found.

With our non-detection and an accurate quantification of the recovery efficiency of our pipeline, we set 95 per cent confidence upper limits on the rate of failed supernovae. We suggest that failed supernovae are likely to be no brighter than an absolute magnitude of -11. We compare our calculated rates of failed supernovae to the rate of core-collapse supernovae in the same data set, and find upper limits on the ratio of these rates. While these rates are not very restrictive, they do represent the most constraining observational limits thus far.

Further long-term observations of nearby galaxies, in particular with the Vera C. Rubin Observatory will allow us to study these events in greater detail.

\section*{Acknowledgements} \label{sec:acknowledgements}


We thank Se\'an Brennan for advice on PSF injection, as well as Emma Callis, Shane Moran, and Beth Fitzpatrick for helpful discussions regarding the project. We thank John Beacom for suggesting Figure \ref{fig:exclusion}, and we thank the anonymous referee for their comments and suggestions.

The research conducted in this publication was funded by the Irish Research Council under grant number GOIPG/2020/542. MF acknowledges support from a Royal Society - Science Foundation Ireland University Research Fellowship. 

This research has made use of the NASA/IPAC Extragalactic Database (NED), which is operated by the Jet Propulsion Laboratory, California Institute of Technology, under contract with the National Aeronautics and Space Administration. 

This paper is based on observations obtained with the Samuel Oschin Telescope and the 60-inch Telescope at the Palomar Observatory as part of the Palomar Transient Factory project, a scientific collaboration between the California Institute of Technology, Columbia University, Las Cumbres Observatory, the Lawrence Berkeley National Laboratory, the National Energy Research Scientific Computing Center, the University of Oxford, and the Weizmann Institute of Science. 

Based on observations obtained with the Samuel Oschin 48-inch Telescope at the Palomar Observatory as part of the Zwicky Transient Facility project. ZTF is supported by the National Science Foundation under Grant No. AST-1440341 and a collaboration including Caltech, IPAC, the Weizmann Institute for Science, the Oskar Klein Center at Stockholm University, the University of Maryland, the University of Washington, Deutsches Elektronen-Synchrotron and Humboldt University, Los Alamos National Laboratories, the TANGO Consortium of Taiwan, the University of Wisconsin at Milwaukee, and Lawrence Berkeley National Laboratories. Operations are conducted by COO, IPAC, and UW. 

This research is based on observations made with the NASA/ESA Hubble Space Telescope obtained from the Space Telescope Science Institute, which is operated by the Association of Universities for Research in Astronomy, Inc., under NASA contract NAS 5–26555. These observations are associated with program(s) 9073, 10452, 15645, and 16508. 

This research made use of Astropy\footnote{\url{http://www.astropy.org}}, a community-developed core Python package for Astronomy \citep{astropy:2013, astropy:2018}. 

This work made use of v2.2.1 of the Binary Population and Spectral Synthesis (BPASS) models as described in \cite{bpass} and \cite{bpass2}.

\section*{Data Availability}

The PTF and ZTF survey data used in this study are available from \url{https://irsa.ipac.caltech.edu/frontpage/}.








%
%
%
%
%
\bsp	
\label{lastpage}
\end{document}